\documentclass{article}

\usepackage{PRIMEarxiv}

\usepackage[utf8]{inputenc} % allow utf-8 input
\usepackage[T1]{fontenc}    % use 8-bit T1 fonts
\usepackage{hyperref}       % hyperlinks
\usepackage{url}            % simple URL typesetting
\usepackage{booktabs}       % professional-quality tables
\usepackage{amsfonts}       % blackboard math symbols
\usepackage{nicefrac}       % compact symbols for 1/2, etc.
\usepackage{amsmath}
\usepackage{siunitx}
\usepackage{microtype}      % microtypography
\usepackage{lipsum}
\usepackage{fancyhdr}       % header
\usepackage{graphicx}       % graphics
\graphicspath{{media/}}     % organize your images and other figures under media/ folder
\usepackage{inconsolata}
\newcommand{\x}{{\mathbf{x}}}
\DeclareMathOperator*{\argmin}{argmin}

\usepackage[export]{adjustbox} 

\usepackage[labelfont=bf, labelsep=period]{caption}
\usepackage{multirow}
\usepackage{float}

\title{CRYSIM: Prediction of Symmetric Structures of Large Crystals with GPU-based Ising Machines
}

\author{
	\textbf{Chen Liang}\textit{$^{1}$}, \textbf{Diptesh Das}\textit{$^{1}$}, \textbf{Jiang Guo}\textit{$^{1}$}, \textbf{Ryo Tamura}\textit{$^{1, 2}$}, \textbf{Zetian Mao}\textit{$^{1}$}$^{\ast}$, \textbf{Koji Tsuda}\textit{$^{1, 2, 3}$}$^{\ast}$ \\ \\
	$^{1}$~Department of Computational Biology and Medical Sciences, Graduate School of Frontier Sciences, \\ The University of Tokyo, 5-1-5 Kashiwanoha, Kashiwa 277-8561, Japan. \\
	$^{2}$~Center for Basic Research on Materials, National Institute for Materials Science, 1-1 Namiki, \\ Tsukuba, Ibaraki 305-0044, Japan.\\
	$^{3}$~RIKEN Center for Advanced Intelligence Project, 1-4-1 Nihonbashi, Chuo-ku, Tokyo 103-0027, Japan.\\
	$^\ast$~Corresponding email: \url{zt.mao97@gmail.com}, \url{tsuda@k.u-tokyo.ac.jp} 
}

\begin{document}

\maketitle

\begin{abstract}
  Solving black-box optimization problems with Ising machines is increasingly common in materials science.
  However, their application to crystal structure prediction (CSP) is still ineffective
  due to symmetry agnostic encoding of atomic coordinates.
%  Our algorithm called CRYSIM encodes the space group, the Wycoff position combination and the coordinates of independent sites separately.
  We introduce CRYSIM, an algorithm that encodes the space group, the Wyckoff positions combination, and coordinates of independent atomic sites as separate variables.
  This encoding reduces the search space substantially by exploiting the symmetry in space groups.
  When CRYSIM is interfaced to Fixstars Amplify, a GPU-based Ising machine, 
  its prediction performance was competitive with CALYPSO and Bayesian optimization
  for crystals containing more than 150 atoms in a unit cell.
  Although it is not realistic to interface CRYSIM to current small-scale quantum devices,
  it has the potential to become the standard CSP algorithm in the coming quantum age. 
\end{abstract}

% keywords can be removed

\section{Introduction}
\label{sec:Introduction}
The advancement of various significant technology fields relies on discovery of innovative materials with desired chemical or physical properties \cite{RN1145}, and obtaining correct structures of materials, arrangement of atoms in the unit cell, is the prerequisite.
To achieve the goal, crystal structure prediction (CSP) \cite{RN682, RN1147}, in which the most stable crystal structure is inferred only from its chemical composition, has been widely adopted.
%In crystal structure prediction (CSP),
%the most stable crystal structure is inferred only from its chemical composition.
The vast configuration space and the richness of local minima
on potential energy surfaces (PESs)
renders CSP a challenging task~\cite{RN959}.
Optimization algorithms,
such as genetic algorithms~\cite{RN969, RN971, RN1137, RN950, RN681},
particle-swarm optimization~\cite{RN949, RN951},
Bayesian optimization (BO)~\cite{RN686, RN958, doi:10.1080/27660400.2022.2055987},
are proposed and successfully applied in practice.
Typically, they create roughly-shaped initial structures and the final optimization is done by a geometric relaxation software
either based on first-principles calculation or pretrained universal neural network potentials (NNPs). 
Nevertheless, these methods generally require a great number of iterations.
In recent years, deep learning-based crystal generative models \cite{xie2022crystal, RN974, jiao2024space, RN1160, RN1167, RN1159, RN1157} are developing fast,
but they might find problems in extrapolation outside their training datasets.
Therefore, as an example, due to the scarcity of corresponding data, CSP on 2D materials \cite{RN1136, RN1135} and nanoclusters \cite{RN1134, RN975, RN1146} generally relies on optimization methods.
Besides, in both categories, most of the methods work well for crystals
containing less than 60 atoms \cite{RN1160} in a unit cell,
but are still not ideal for larger crystals.
For example, the training data of CDVAE \cite{xie2022crystal} and MatterGen \cite{RN1157} does not include
large crystals with more than 20 atoms in a unit cell. 
GNoME \cite{RN974} successfully explores larger ones approximating 100 atoms, but considerable computational cost is required.

Ising machines~\cite{RN1107, RN1151} are hardware-assisted discrete optimizers that solve a quadratic unconstrained binary optimization (QUBO) problem,
\begin{equation} \label{eq:qubo}
  \x^* = \argmin_{\x \in \{0,1\}^M} \sum_{i=1}^M h_i x_i + \sum_{i,j=1}^M
  J_{ij} x_i x_j,   
\end{equation}
where $\x$ is an $M$-dimensional bit vector and
$h_i$ and $J_{ij}$ are real-valued parameters.
CSP can be represented as a QUBO problem,
either by simplifying the energy function~\cite{RN965,RN1141,RN988}
or the use of a surrogate machine learning model~\cite{RN1155, RN1166}.
Among the existing studies, 
Gusev et al. \cite{RN965}, Ichikawa et al. \cite{RN1141} and Xu et al. \cite{RN1155} provided
QUBO formulations of CSP under a fixed space group.
Couzinié et al. \cite{RN988} and Couzinié et al. \cite{RN1166} disregarded the space group and
employed a grid-based representation of atomic coordinates
in their QUBO formulation.
%Xu et al.\cite{RN1155} integrate an activate learning to
%optimize the lattice parameters only.}
Notably, they lack a feature of dynamically adjusting the space group,
which is common in state-of-the-art CSP algorithms such as CALYPSO \cite{RN949, RN951}, USPEX \cite{RN1137, RN681, LYAKHOV20131172}
and CRYSPY \cite{RN958, doi:10.1080/27660400.2022.2055987}.
Futhermore, these algorithms have not been tested on complex problems
mainly due to the scale restriction of D-Wave \cite{8728085} quantum annealer.

In this work, we develop a method named CRYSIM
(CRYstal structure prediction  with Symmetry-encoded Ising Machine).
Our bit vector represents the lattice parameters, the symmetry information
including the crystal system (CS), the space group (SG) and the Wycoff positions combination (WPC), and coordinates of independent sites.
This bit vector is translated to a crystal structure and M3GNet \cite{RN1130} provides its potential energy.
Our goal is to find the optimal bit vector that gives the lowest potential energy.
To enable the search with an Ising machine, a factorization machine (FM)~\cite{5694074, RN1110, guo2024boostingqualityfactortamm, RN1154} is trained
with available pairs of bit vectors and corresponding energies with an active learning workflow \cite{RN685}.
Since the prediction function of an FM is quadratic,
the optimal bit vector that minimizes the FM-approximated potential energy
can be found with an Ising machine.
It does not always coincide with the real optimal solution,
but one can expect that the error decreases as the amount of training data
increases during the search process. 
Our information-rich bit vector inevitably inhibits the use of D-Wave quantum annealers.
Instead, we employ a GPU-based Ising machine, Fixstars Amplify \cite{RN1120}, to solve a problem with over several thousand bits.
It is based on simulated annealing and uses multi-level parallel
processing on multiple GPUs to find optimal solutions.
Fixstars Amplify relies on conventional semiconductor technologies,
but can handle large scale problems up to 130,000 bits with full connection.
It has been employed in molecular generation \cite{RN964}, materials design \cite{RN1161, RN1162, RN1163} and various engineering fields \cite{RN1165, RN1164}.

Our method outperforms BO \cite{RN686}
and CALYPSO \cite{RN949, RN951} on three small crystals
as well as large ones containing more than 150 atoms in unit cells.
Notably, CRYSIM is the only model that successfully generates the ground truth
Ca$_{24}$Al$_{16}$(SiO$_4$)$_{24}$ structure,
containing 160 atoms in the unit cell, within 300 relaxations.
In this work, GPUs are adopted, but 
CRYSIM can leverage any Ising machines including rapidly developing quantum devices.

\section{Results}
\label{sec:Results}

\subsection{Bit Vector Encoding}
The binary representation in CRYSIM consists of the following three parts: lattice parameters,
symmetry information and
3D coordinates of independent sites. 
%\textcolor{red}{In order to represent a crystal consisting of $k$ element species, the lattice is discretized into a $g \times g \times g$ grid }
%denoted by the number of atom species $k$,
%and by the discretization resolution $g$.
In the first part, the six dimensional lattice parameters
are individually discretized
and summarized into a bit vector with one-hot encoding.
The second part includes a crystal system (CS), a space group (SG), a group of Wycoff positions combinations (WPCs). Sizes of each vector segment depend on the set of
all possible space groups compatible with the given chemical composition, which is determined by whether there exists at least one WPC for achieving symmetry of the SG. Similarly, only compatible CSs are included in the embeddings. Accordingly, 
if $m$ crystal systems are involved,
each of which has $s_1,\ldots,s_m$ compatible space groups,
the CS part has $m$ bits to represent the CS
and the SG part has $\max_{i=1,\ldots,m} s_i$ bits to represent
the SG. 
If the crystal structure has the $i$-th CS and $j$-th SG, the corresponding bits are set as 1 and the remaining are 0s.
Given the SG, 30,000 plausible WPCs are generated and sorted in descending order based on the maximum multiplicity of involved WPs, so that
%To this aim, we engineered the WP combination generator in GN-OA package\cite{RN678}
more plausible combinations are prioritized \cite{LYAKHOV20131172, RN957}.
The WPCs are divided into 300 groups of the size 100, which is encoded in the WPC segment with a 300-dimensional one-hot vector for specifying the group.
We engineered the WPC generator in GN-OA package \cite{RN678} to derive the set of compatible SGs by computing comprehensive lists of WPCs according to the input chemical composition for all SGs \cite{Deng:db5062,AVERY2017208}.
The third part consists of $k$ copies of a $g^3$-dimensional bit vector in order to represent a crystal containing $k$ element species, in which $g$ denotes lattice discretization resolution (LDR).
A 3D $g \times g \times g$ grid is assumed within the unit cell.
If an independent site of the atom species
exists near a grid point, the corresponding bit is set to 1.
In decoding, 100 structures are generated corresponding to
all WPCs in the specified WPC group.
Among them, the one with the largest minimum interatomic distance (MID)
is selected to increase the possibility of deriving stable states \cite{RN959}. Details of the encoding and decoding procedures are provided in \textbf{Method}.
Besides, \textbf{Section \ref{sec:wpc}} and \textbf{\ref{sec:discuss}} of \textbf{Supplementary Information} presents a detailed explanation about WPCs generation and application.

\subsection{CRYSIM Workflow}

\begin{figure}[htbp]
	\centering
	\includegraphics[width=\linewidth, center]{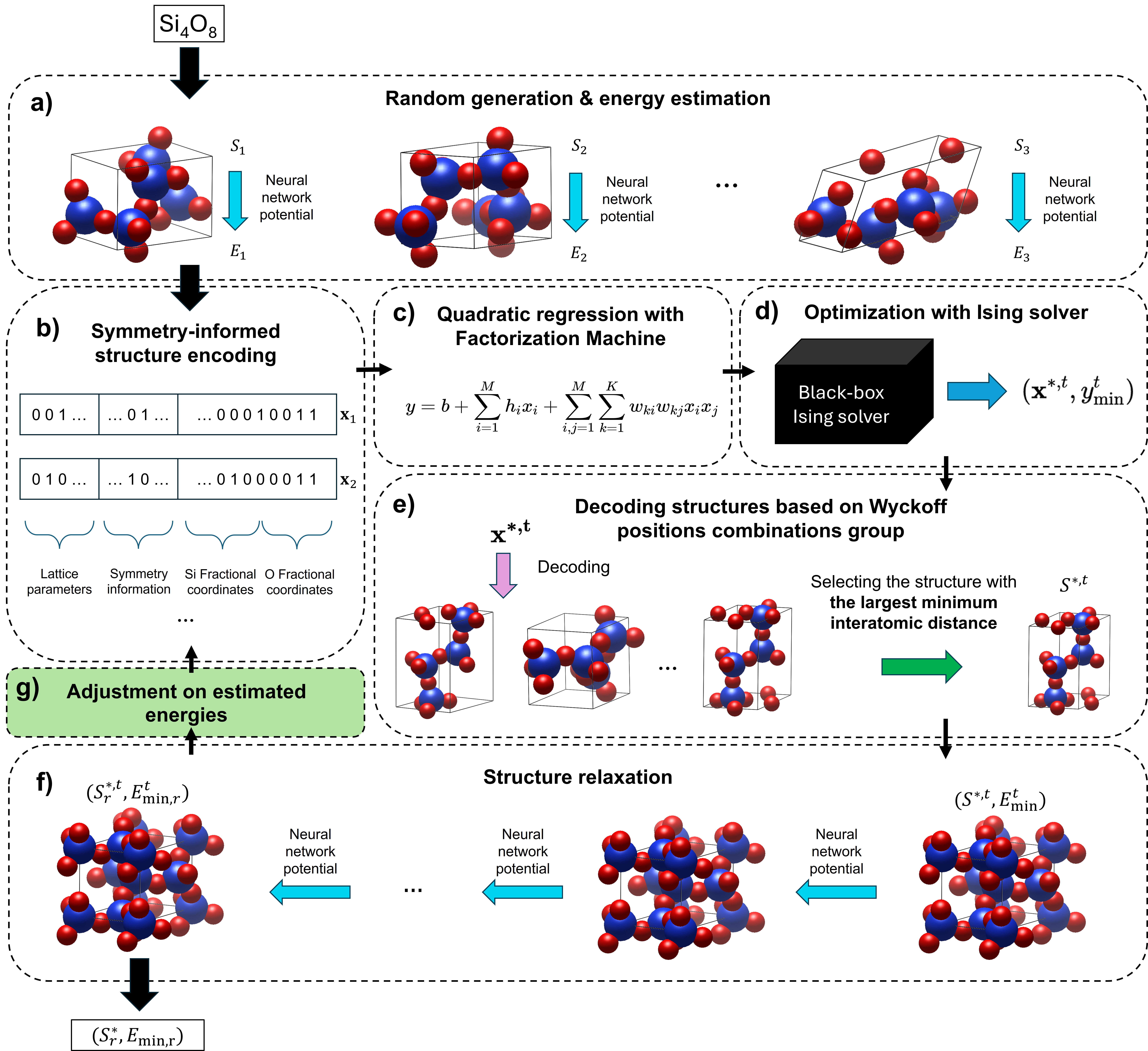}
	\caption{The workflow of CRYSIM that contains $T$ iterations, using Si${_4}$O${_8}$ as an illustration. Thin arrows denote the workflow at the $t$-th iteration, and thick arrows denote entering and exiting iterations.
		\textbf{a} Given the considered material system, a dataset is obtained by RG to provide training samples and determine the upper bound of lattice parameters for binary representation. Potential energy of each material is also estimated by pretrained NNP without structure relaxation.
		\textbf{b} Structures in the dataset $\{S_1, S_2, \ldots, S_{1000}\}$ are encoded into binary vectors $\{\mathbf{x}_1, \mathbf{x}_2, \ldots, \mathbf{x}_{1000}\}$ using symmetry-informed integer encoding.
		\textbf{c} FM is used to perform regression from the binary vectors to their corresponding estimated energies, obtaining the objective function to be optimized.
		\textbf{d} An Ising solver is employed to solve the learned objective function to minimize $y$ in $t$-th iteration, resulting in $\mathbf{x}^{*,t}$. Amplify is used in this work.
		\textbf{e} The solved binary embeddings $\mathbf{x}^{*,t}$ is decoded into crystal structures. Since one bit in the WPC segment represents a group of 100 WPCs, 100 structures are derived. The one with the largest MID is selected as $S^{*,t}$. We note that the Si$_4$O$_8$ structures drawn in the figure \textbf{e} are indicative, which have different SGs.
		\textbf{f} The solved structure $S^{*,t}$ is relaxed by NNP, leading to a structure-energy pair $(S^{*,t}_r, E^t_{\min, r})$.
		If iterations have not finished, frames in the relaxation trajectory are sampled.
		\textbf{g} Among the sampled structures, if one contains an MID smaller than 0.5 \AA\ but still is estimated to have a negative energy, the energy is reassigned with a high positive one before adding the points into the training dataset for the next iteration.
		After finishing all iterations, the final structure $S_{r}^{*}$, the one with the lowest relaxed energy among all crystals in all $T$ iterations, will be regarded as the discovered stable structure of this system.}
	\label{fig:workflow}
\end{figure}

The workflow of CRYSIM is depicted in \textbf{Fig. \ref{fig:workflow}}. 
First, 1000 initial structures are obtained by random generation (RG) developed in this work (see \textbf{Method} for details)
with the given chemical composition,
and converted to bit vectors $\x_l$.
Their potential energies $y_l$ are estimated
using M3GNet without structure relaxation.
The training dataset is described as the pairs of bit vectors
and energies, i.e., $D = \{(\x_l, y_l)|l=1, 2, \ldots, 1000\}$, which is then used to train an FM model ~\cite{5694074, RN1110}.
The functional form of FM is described as
\begin{equation}
y = b + \sum_{i=1}^M h_i x_i + \sum_{i,j=1}^M
\sum_{k=1}^K w_{ki} w_{kj} x_i x_j,
\end{equation}
where $b$, $h_i$ and $w_{ki}$ are real-valued parameters, and $x_i$ is the $i$-th element of a vector $\x$.
It is similar to QUBO, but the weight matrix of
quadratic terms is a low-rank matrix parameterized by $w_{ki}$ and $w_{kj}$.
Then, Fixstars Amplify \cite{RN1120} is used to optimize the bit vector to
minimize the energy approximated by FM.
Based on the solution, 100 structures with the same SG but different WPCs from the solved WPCs group are translated back to crystal structures, and the one with the largest MID is selected and relaxed using M3GNet.
After relaxation, we sample 30 structure frames from the relaxation trajectory. Among the samples, if a structure has an MID lower than 0.5 \AA\ but is still assigned a negative energy, the energy is adjusted to a high positive value to mitigate negative impact due to inaccuracy of NNP. Then the data points are added to
$D$ and FM is retrained.
The above procedure is repeated $T=300$ times and
the most stable structure is recorded as the final result.

\subsection{Recovering Benchmark Materials}
\begin{figure}[t]
	\centering
	\includegraphics[width=0.8\linewidth, center]{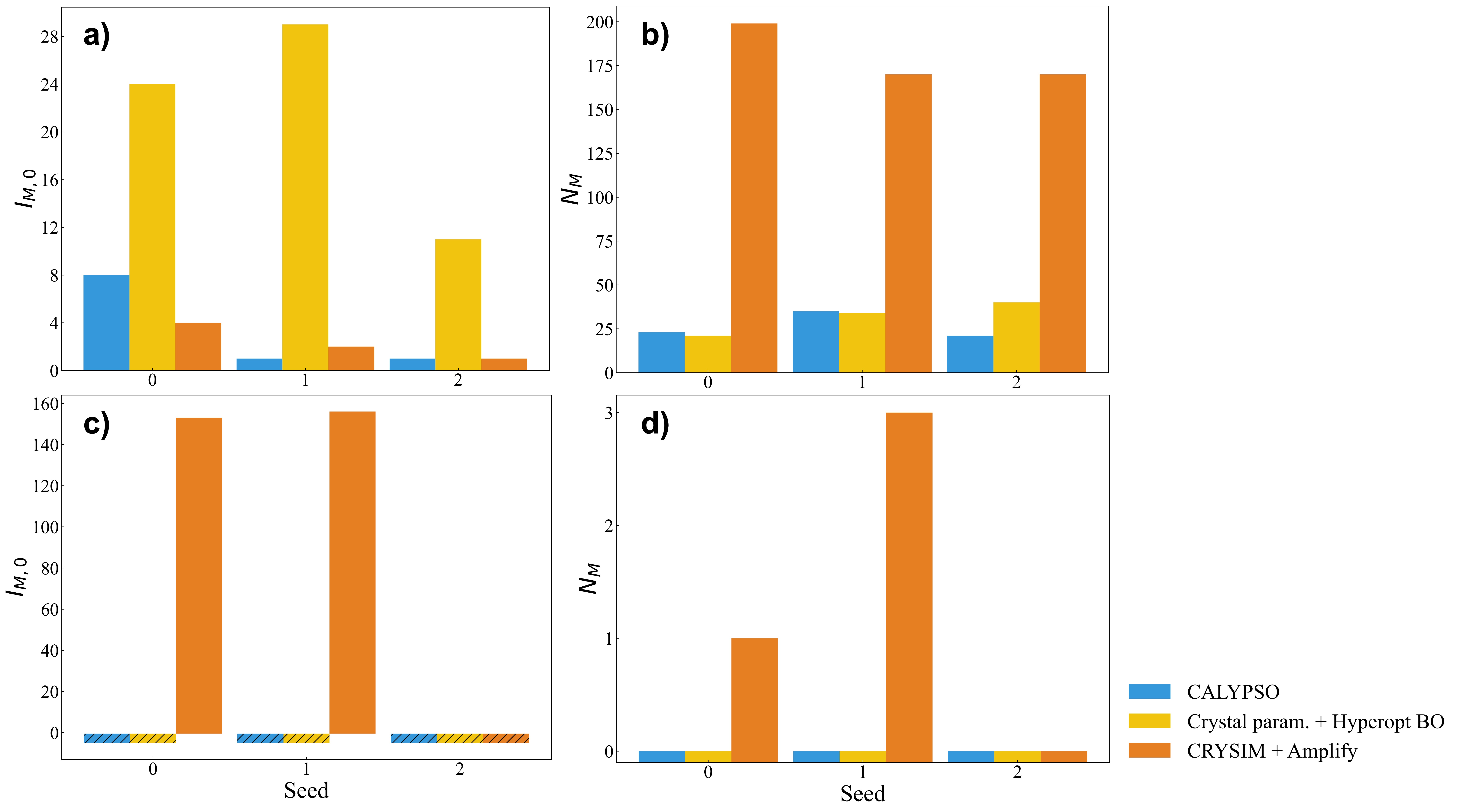}
	\caption{The first iteration when the generated structure matches the ground truth ($I_{M, 0}$), and the number of successfully matched structures among the generated ones with the lowest energy ($N_M$) of \textbf{a-b} Ca$_4$S$_4$ and \textbf{c-d} Ba$_3$Na$_3$Bi$_3$ for the three optimization methods in 300 iterations.
	Shadowed bars in \textbf{c} indicate that the corresponding methods fail to find the ground truth structure with these seeds.
	%In the legends, "Crystal param. + Hyperopt BO" represents BO that optimizes crystal parameters implemented based on hyperopt.
	%The number after "CRYSIM" indicates LDR used for dividing lattices.
	}
	\label{fig:benchbar}
\end{figure}

\begin{figure}[t]
	\centering
	\includegraphics[width=\linewidth, center]{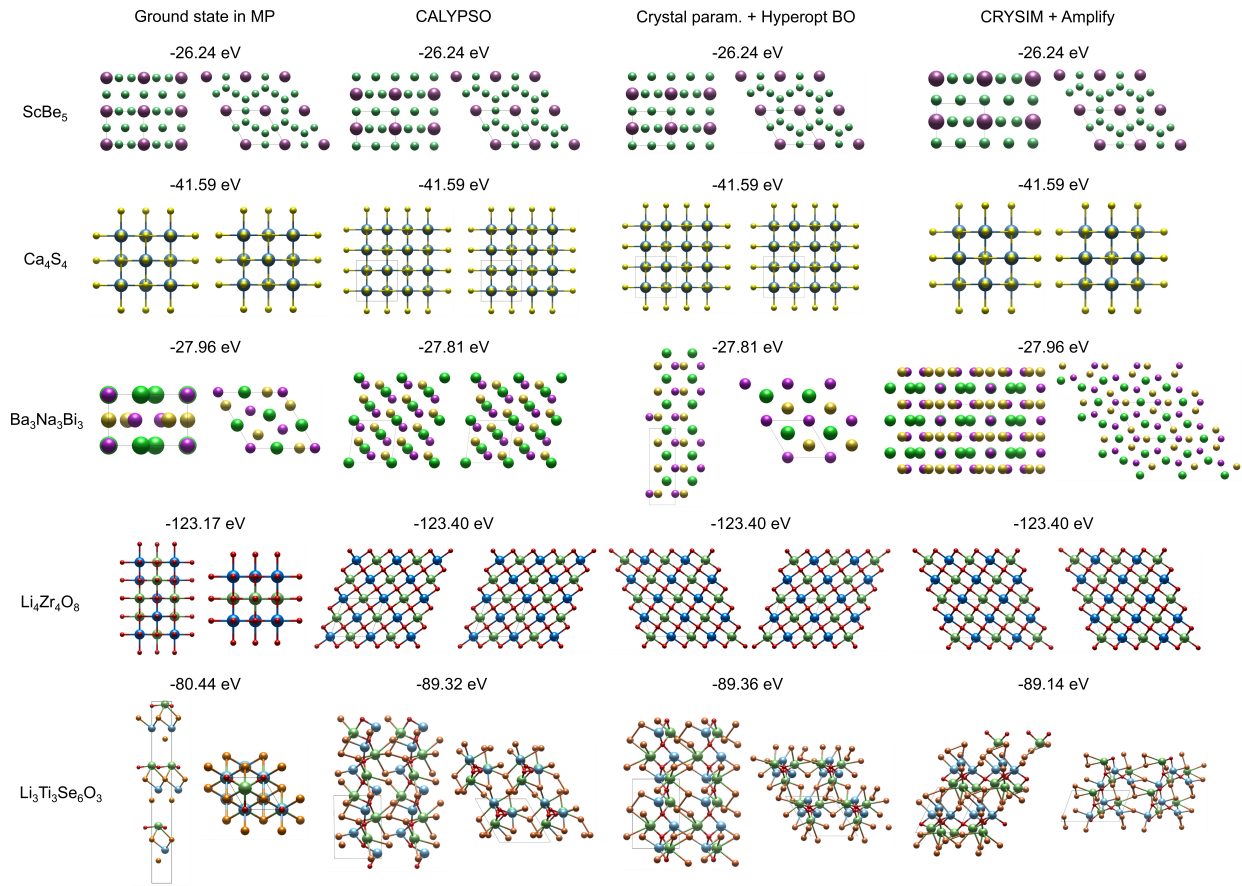}
	\caption{Side view (left column for each method) and top view (right column) of ground states in MP of the five benchmark crystals, mp-11277 (Sc: purple, Be: green), mp-1672 (Ca: blue, S: yellow), mp-31235 (Ba: green, Na: yellow, Bi: pink), mp-755253 (Li: green, Zr: blue, O: red), and mp-1211008 (Li: green, Ti: blue, Se: orange, O: red), respectively, and predicted configurations by three CSP methods after structure relaxation, visualized by VESTA software \cite{Momma:db5098}, with M3GNet-estimated relaxed energies labeled above.
		%In the legends, "Crystal param. + Hyperopt BO" represents BO that optimizes crystal parameters implemented based on hyperopt.
		%The number after "CRYSIM" indicates LDR used for dividing lattices.
		Most configurations are expanded into superlattices to display the patterns. Crystals with the lowest energies are selected. If there are more than one crystals having the same energy, the one obtained in the earliest iteration is shown.}
	\label{fig:bench-fig}
\end{figure}

We begin with relatively simple benchmark crystal tasks to demonstrate CRYSIM's ability to address CSP.
Wei et al. \cite{RN989} developed a series of quantity measurements for evaluating CSP algorithms, and selected five crystals, including ScBe$_5$, Ca$_4$S$_4$, Ba$_3$Na$_3$Bi$_3$, Li$_4$Zr$_4$O$_8$ and Li$_3$Ti$_3$Se$_6$O$_3$, as examples to conduct tests.
Ground states of all compounds are determined based on the Materials Project (MP) database \cite{RN743}, i.e., mp-11277, mp-1672, mp-31235, mp-755253 and mp-1211008, respectively.
%Following their work, BO, CALYPSO and CRYSIM are employed to predict the five crystal structures.
%In this section, benchmark crystals are adopted to demonstrate the ability of CRYSIM to address CSP.
Classical algorithms considered for comparison include
%simple RG (see \textbf{Method}), PyXtal \cite{RN957}-based RG in CRYSPY \cite{RN958},
CALYPSO \cite{RN949, RN951} and simple BO that directly optimizes lattice parameters, fractional coordinates, the SG number and the WPCs index of crystals \cite{RN678}, implemented based on the hyperopt package \cite{osti_10022769}, denoted as "Crystal param. + Hyperopt BO".
All methods are limited to perform 300 times of structure relaxation during one run to make a fair comparison. Accordingly, CALYPSO is leveraged for 30 generations, with the population size per iteration set as 10.
Besides, in all experiments in this article, structures containing interatomic distances smaller than 1.0 \AA\ are excluded from the statistics, to ensure that all crystals remain physically valid.
Tests of each method repeat three times with different seeds.
%For CRYSIM, the LDR $g$ is set as 12, denoted as "CRYSIM-12".
Training settings of FM, values of hyperparameters for Amplify and classical algorithms are reported in \textbf{Section \ref{sec:ising}} and \textbf{\ref{sec:hp}} of \textbf{Supplementary Information}.
%Hyperparameters settings are the same as previous experiments.

We use \texttt{StructureMatcher} function from the \texttt{pymatgen} package \cite{ONG2013314} to
%decide whether a predicted material is comparable with its ground truth.
determine structural similarity between predicted and known ground truth materials.
The function can compute minimum average pair-wise displacement between two corresponding atoms in two configurations among all permutations.
The predicted structure successfully matches the ground truth as long as the displacement is computable, which suggests that \texttt{StructureMatcher} is able to distinguish corresponding atoms between them.
Details of criterion of matching is provided in \textbf{Method} section.

When assessing results, we only include structures reaching the lowest energy ($E_{\min}$) among all obtained ones in 300 iterations, i.e.,
%all structures with energies within the range
%$\{E|E_{\min} < E < E_{\min} + 0.02\ \mathrm{eV}\}$
$\{E|E = E_{\min}\}$,
and compare them against the ground truth.
%where $E_{\min}$ is the lowest energy among all obtained structures in 300 iterations.
Several major metrics are defined for this task:
(1) $I_{M, 0}$ denotes the first iteration at which the ground truth is identified;
%(2) $E_{\min}$ denotes the lowest energy among all identified correct structures.
(2) $N_E$ denotes the number of iterations reaching the lowest energy, i.e.,
%$N_E=|\{E|E_{\min} < E < E_{\min} + 0.02\ \mathrm{eV}\}|$.
$N_E=|\{E|E = E_{\min}\}|$;
(3) $N_M$ denotes the number of successfully matched ones among all considered structures.
We provide a further illustration on evaluation of CSP algorithms in \textbf{Discussion} section.
\textbf{Fig. \ref{fig:benchbar}} summarizes results of two representative crystals, and comprehensive information is presented in \textbf{Table \ref{table:bench-scbe5}-\ref{table:bench-litise2o}}.
Other metrics, such as displacement calculated by \texttt{StructureMatcher} of the structure in iteration $I_{M, 0}$, denoted as $D_{M, 0}$, as well as the minimum displacement $D_{M, \min}$ and corresponding iteration $I_{M, \min}$, are also reported.
%Besides, we claim that the 0.02 eV screening threshold is appropriate, since almost all selected materials also match with the stable states ($N_E=N_M$) in successful runs.
Predicted configurations with the lowest estimated relaxed energies in the three trials are shown in \textbf{Fig. \ref{fig:bench-fig}}.

Ground states of ScBe$_5$ and Ca$_4$S$_4$ can be readily discovered by all three methods,
%The ScBe$_5$ task is especially simple for CALYPSO, which is accomplished within the 10 initial materials in the first cycle in all three trials.
but CRYSIM generates significantly more stable states than the two classical methods. Besides, smaller $I_{M, 0}$ of CRYSIM optimizers indicate that FM can quickly and effectively characterize the PES by learning from initial datasets.
The superiority of CRYSIM becomes notable for the more complicated Ba$_3$Na$_3$Bi$_3$ system, in which CRYSIM is the only method that successfully discovers the stable state with correct estimated energies.

All methods fail on Li$_4$Zr$_4$O$_8$ and Li$_3$Ti$_3$Se$_6$O$_3$ structure prediction if only the crystals of $E_{\min}$ are counted, which, however, reach a even lower estimated relaxed energy than the ground states.
%in which CALYPSO fails in all three trials, CRYSIM-empowered BO still exhibits a larger possibility to find the ground truth much faster than simple BO, with an 83 first-found index versus 264 and 10 versus 192 in the two successful trials. Li$_3$Ti$_3$Se$_6$O$_3$ is the only crystal that all methods fail. 
This may be attributed to two main reasons.
First, the selected benchmark structures may not represent the ground states of corresponding chemical compositions, suggested by positive energies above hull.
Especially, Li$_3$Ti$_3$Se$_6$O$_3$ (mp-1211008) exhibits a substantial energy above hull, 0.617 eV/atom according to MP, indicating a potential to transform into alternative phases predicted by the optimization methods.
Second, the pretrained NNP leveraged in this study is not sufficiently accurate, and a 0.01 eV variation in potential energy can affect the behavior of optimization algorithms.
As an instance, we further introduce predicting results on Li$_8$Zr$_4$O$_{12}$ (mp-4156), a stable structure of Li-Zr-O family that has been observed in experiments, in \textbf{Fig. \ref{fig:li8zr4o12}} and \textbf{Table \ref{table:li8zr4o12}}.
In all configurations with the lowest energies, hexa-atomic rings absent in mp-4156 can be found, which may suggest an intrinsic bias of the NNP on energy estimation.

\subsection{Large Crystal Structures Prediction}

\begin{figure}[t!]
	\centering
	\includegraphics[width=\linewidth, center]{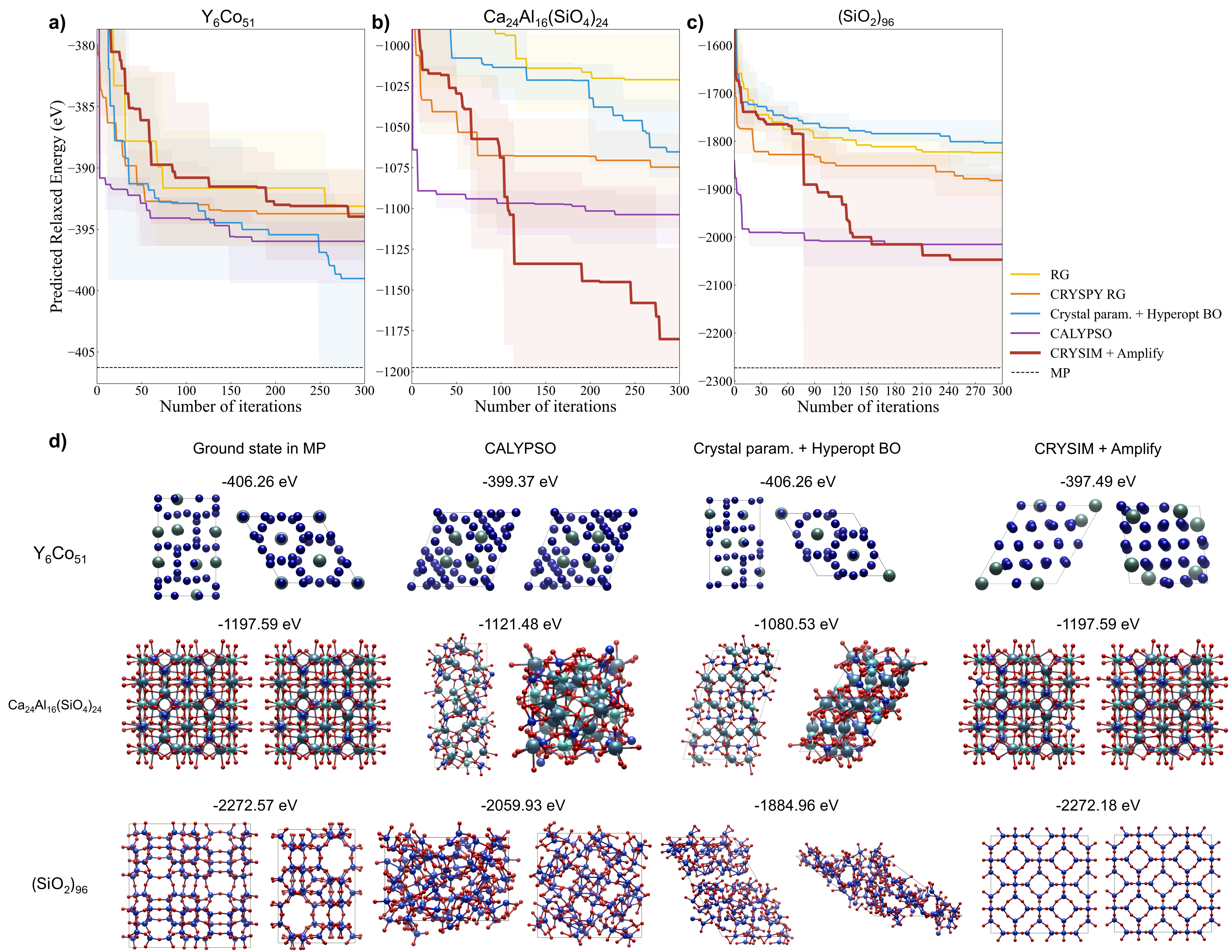}
	\caption{Averaged accumulated lowest M3GNet-estimated relaxed energies of \textbf{a} Y$_6$Co$_{51}$, \textbf{b} Ca$_{24}$Al$_{16}$(SiO$_4$)$_{24}$ and \textbf{c} (SiO$_2$)$_{96}$ structures derived from various CSP algorithms.
		% baseline algorithms, including RG, RG in CRYSPY, simple BO and CALYPSO, are drawn with silver, light grey, deep grey and black lines
		%In the legends, "CRYSPY RG" and "Crystal param. + Hyperopt BO" represent PyXtal-aided RG implemented in CRYSPY and BO that optimizes crystal parameters implemented based on hyperopt, respectively.
		%The numbers after "CRYSIM" indicate LDRs used for dividing lattices.
		%CRYSIM-based methods with Amplify as solvers are represented by red curves, and deeper colors suggest higher LSNs.
		Each curve is averaged on five tests with different random seeds, and colored shaded areas cover the maximum and minimum in the five trials.
		Dash lines are relaxed energies of ground truth materials in MP.
		\textbf{d} Side view (left column for each method) and top view (right column) of ground states in MP, mp-1106140 (Y: grey, Co: blue), mp-6008 (Ca: grey, Al: light blue, Si: deep blue, O: red), and mp-1200292 (Si: blue, O: red), and representative predicted configurations after structure relaxation, respectively, visualized by VESTA software \cite{Momma:db5098}, with M3GNet-estimated relaxed energies labeled above.}
	\label{fig:main}
\end{figure}

This section introduces experiment results on three large material systems, including Y$_6$Co$_{51}$, Ca$_{24}$Al$_{16}$(SiO$_4$)$_{24}$ and (SiO$_2$)$_{96}$, to demonstrate capability of CRYSIM.
The crystal Y$_2$Co$_{17}$ has been a classical benchmark for assessing CSP algorithms \cite{RN686, RN1152, PhysRevMaterials.5.054408}, but the two stable structures in this material family recorded in MP can only be achieved with unit cells Y$_4$Co$_{34}$ (mp-570718) and Y$_6$Co$_{51}$ (mp-1106140). Since the number of atoms in unit cells are not optimized in CRYSIM, we start directly with Y$_6$Co$_{51}$.
Ca$_{24}$Al$_{16}$(SiO$_4$)$_{24}$ (mp-6008) \cite{RN965, LYAKHOV20101623} and (SiO$_2$)$_{32}$ \cite{LYAKHOV20101623} have also been discussed in previous works as examples of CSP on large crystals. Here, (SiO$_2$)$_{96}$ (mp-1200292) is chosen since it is the largest SiO$_2$ crystal in MP that has been observed in experiments.

Apart from CALYPSO and BO introduced earlier, simple RG, which is employed to generate initial training set for CRYSIM, and PyXtal \cite{RN957}-based RG in CRYSPY \cite{RN958}, denoted as "CRYSPY RG", are additionally included as baseline CSP methods.
300 times of structure relaxation, i.e., 300 iterations, are conducted in one run, and tests of each method repeat five times with different seeds.
%LDR for CRYSIM are set as 12, 12 and 15 for Y$_6$Co$_{51}$, Ca$_{24}$Al$_{16}$(SiO$_4$)$_{24}$ and (SiO$_2$)$_{96}$, respectively, represented as "CRYSIM-12" and "CRYSIM-15", respectively.
\textbf{Fig. \ref{fig:main}a-c} presents averaged accumulated lowest energies of crystals during generation in the 300 cycles.
Optimal materials found by each algorithm are visualized in \textbf{Fig. \ref{fig:main}d}. 
Corresponding data is summarized in \textbf{Table \ref{table:mainresults}},
as well as \textbf{Table \ref{table:y6co51}-\ref{table:si96o192}} for metrics defined in the last section.
%The three tasks are more difficult than the benchmark crystals, and therefore metrics defined in the last section do not effectively reflect difference in optimization power.

%Besides, we use \texttt{StructureMatcher} function in \texttt{pymatgen} package \cite{ONG2013314} to decide whether a predicted material is comparable with its ground truth, in which parameters are set as \texttt{stol=0.5},  \texttt{ltol=0.3},  \texttt{angle\_tol=10.0}, the same as other related works \cite{jiao2023crystal, sriram2024flowllm}.
%This function will calculate minimum normalized average root mean square pair-wise displacement between two input structures among all atom permutations. But if corresponding atoms in the two structures are not detected, which means that the function cannot find any similarity between them, the calculation will not be proceeded.
%Based on that, representative generated materials are exhibited in \textbf{Fig. \ref{fig:main}d}.

%If the ground truth in MP is found, i.e., the pair-wise displacement can be calculated, the corresponding generated configuration is provided. Otherwise, we show the one with the lowest energy. 

On Y$_6$Co$_{51}$, BO is the only method that discovers the ground state with a -406.26 eV energy, the same as corresponding relaxed energy of mp-1106140. 
However, computational complexity of BO scales with the total number of atoms \cite{frazier2018tutorialbayesianoptimization, RN1156, pmlr-v161-eriksson21a}, leading to a significantly reduced performance on Ca$_{24}$Al$_{16}$(SiO$_4$)$_{24}$ and (SiO$_2$)$_{96}$, even falling below CRYSPY RG.
CALYPSO exhibits a higher stability than BO for large systems, and the implementation of pair-wise distance consideration in input renders it unaffected when screening out configurations with small MIDs, as is shown in \textbf{Table \ref{table:filternumclassical}}. However, this feature accelerates the optimization of energies only in the first tens of iterations in \textbf{Fig. \ref{fig:main}b-c}, and then the algorithm is surpassed by CRYSIM methods.

On the other hand, length of CRYSIM embedings is determined solely by the number of elements given the LDR, making it notably advantageous over other algorithms especially on large crystals. 
On the Ca$_{24}$Al$_{16}$(SiO$_4$)$_{24}$ system, which contains 160 atoms in the unit cell, CRYSIM successfully finds the ground truth structure in four out of five trials. For (SiO$_2$)$_{96}$, CRYSIM identifies a configuration with a relaxed energy (-2272.18 eV) close to the stable one (-2272.57 eV), significantly lower than the ones found by other methods.
CSP for large crystals has been a long-standing challenging task. Energy distribution of the configuration space tends to concentrate on unstable states as the system size grows, which means that the difficulty of finding the ground state via RG exponentially increases \cite{RN1092, LYAKHOV20101623, RN1060}. Though CRYSIM does not outperform in all systems, the superiority on large crystals establishes it as a promising approach for CSP.

\subsection{Effects of Processing Techniques}

\begin{table}[t]
	\caption{Lowest energies of structures discovered by CRYSIM optimizers of different LDRs, in which results of integrating MID-related procedures (Y) or absence of it (N) are also shown as an ablation study. \textbf{Bold} values are the lowest average energies for each material system achieved by each LDR, and \underline{underlined} values are the lowest ones of each MID processing strategy among all LDRs. Each value is averaged on three seeds. (unit: eV)}
	\renewcommand\arraystretch{1.5}
	\centering
	\resizebox{16.5cm}{!}{
		\begin{tabular}{ccccccc}
			\hline
			\multirow{2}{*}{System} & \multirow{2}{*}{MID proc.} & \multicolumn{5}{c}{Lattice Discretization Resolution} \\    
			~ & ~ & 5 * 5 * 5 & 7 * 7 * 7 & 9 * 9 * 9 & 12 * 12 * 12 & 15 * 15 * 15 \\
			\hline
			\multirow{2}{*}{Y$_{6}$Co$_{51}$} & N & -388.85±3.83 & -390.59±3.06 & \underline{-392.07±2.19} & -391.16±5.0 & -387.4±0.55 \\
			~ & Y & \textbf{-392.03±0.64} & \textbf{-392.6±4.0} & \textbf{-393.89±1.8} & \underline{\textbf{-394.06±2.15}} & \textbf{-390.72±0.75} \\
			\hline
			\multirow{2}{*}{Ca$_{24}$Al$_{16}$(SiO$_{4}$)$_{24}$} & N & -1067.23±6.05 & -1046.66±11.89 & \underline{-1097.27±12.29} & -1082.55±13.07 & -1064.62±41.43 \\
			~ & Y & \textbf{-1117.44±10.9} & \textbf{-1162.64±16.53} & \textbf{-1142.22±46.27} & \underline{\textbf{-1192.93±4.65}} & \textbf{-1131.17±39.03} \\
			\hline
			\multirow{2}{*}{(SiO$_{2}$)$_{96}$} & N & / & \textbf{-1983.74±31.4} & -1888.92±32.03 & -1874.03±54.39 & \underline{-2050.55±20.04} \\
			~ & Y & / & -1938.85±60.78 & \textbf{-2013.7±80.8} & \textbf{-1940.27±40.23} & \underline{\textbf{-2121.47±109.4}} \\
			\hline
		\end{tabular}
	}
	\label{table:as-pp}
\end{table}

\noindent\textbf{Lattice Discretization Resolution.}\quad
When converting 3D structures into binary vectors, a higher LDR can reduce information loss, nevertheless, leading to exponentially increasing solving difficulty.
%In this section, influence of lattice splits on CRYSIM-based Ising optimizers for material systems of various sizes is discussed.
We investigate the influence of LDR on CRYSIM optimization performance by testing $g\in\{5, 7, 9, 12, 15\}$ across the three large crystals considered in this study, namely Y$_{6}$Co$_{51}$, Ca$_{24}$Al$_{16}$(SiO$_{4}$)$_{24}$ and (SiO$_{2}$)$_{96}$.
\textbf{Table \ref{table:as-pp}} summarizes lowest energies with corresponding average accumulated energy curves recorded in \textbf{Fig. \ref{fig:splits}}, where CRYSIM with different LDRs are denoted as CRYSIM-$g$, such as "CRYSIM-5" for $g=5$.
Each value is the mean of three trials with different random seeds.

Y$_6$Co$_{51}$ and Ca$_{24}$Al$_{16}$(SiO$_4$)$_{24}$ achieve the best CSP results at LDR 12, while (SiO$_2$)$_{96}$ performs the best at 15.
This phenomenon stems from the implementation of CRYSIM, where each atom species is encoded using a $g \times g \times g$ grid of bits, with 0 or 1 indicating the presence of an atom at each discretized unit. Accordingly, a higher LDR is advantageous when the number of atoms per element increases, rather than the total number of atoms.
An LDR of 12 appears to strike a balance between representability and optimization difficulty for Y$_6$Co$_{51}$ and Ca$_{24}$Al$_{16}$(SiO$_4$)$_{24}$, where the maximum number of atoms of a specific element is 51 and 96, respectively. In contrast, (SiO$_2$)$_{96}$, which contains 192 oxygen atoms, may require a higher LDR to better capture interatomic spatial relationship in CRYSIM.
%The three systems have different characteristics from the perspective of system sizes. Y$_6$Co$_{51}$ has two elements and intermediate number of atoms for each species. On the other hand, Ca$_{24}$Al$_{16}$(SiO$_4$)$_{24}$ and (SiO$_2$)$_{96}$ are composed of higher number of elements and atoms per species, respectively, though both of them can be similarly classified into crystals containing large numbers of atoms in total.
%Accordingly, experiment results can unveil the most significant factor with respect to system sizes that influences predicting performance for different optimizers.
%Optimization powers of CRYSIM on Y$_6$Co$_{51}$ and Ca$_{24}$Al$_{16}$(SiO$_4$)$_{24}$ both show a reduction when LDR is increased from 12 to 15, but (SiO$_2$)$_{96}$ reaches the lowest energies with 15. In this regard, we can conclude that a higher LDR is helpful when the number of atoms per element increases, instead of the total number of atoms. The phenomenon is stemmed from the implementation of CRYSIM, in which atoms of one element are encoded in one vector segment.
Numbers of bits for representing each parameter for the three systems are further reported in \textbf{Table \ref{table:length}}.
%Therefore, the embeddings only expand as the number of elements increases for a fixed LDR.

\quad\\
\noindent\textbf{Factorization Machine.}\quad
In this work, FM is employed as the regressor in CRYSIM to build Ising objective functions, and here the fitting accuracy is investigated.
\textbf{Fig. \ref{fig:fm}a-e} show distributions of predicted versus calculated energies of one of the initial Ca$_{24}$Al$_{16}$(SiO$_4$)$_{24}$ datasets derived by RG, the system requiring the largest number of bits to represent due to its chemical composition.
Learnable parameters of FM are decided upon metrics on the validation set, comprised of 10\% of the dataset (see \textbf{Section \ref{sec:ising}} in \textbf{Supplementary Information} for details).
%but all data points are predicted and included in the figures.
Changes of Pearson correlation coefficients (PCCs) and root mean square errors (RMSEs) during training are further provided in \textbf{Fig. \ref{fig:fm}f}.
The consistently high PCC values indicate effective optimization toward the global optimum, despite of fluctuations due to out-of-distribution energies.
Similar trends are observed across all other systems and random seeds.
%Although increase of LDR results in higher difficulty of energy prediction, and the curves may fluctuate due to inclusion of out-of-distribution energies, high PCC values indicate that optimization on the learned objective functions can be effectively proceeded towards the optimum. All the other systems and seeds show similar tendency.

Additionally, a comparison between FM and full-rank quadratic regression (QR), in which quadratic terms in regression functions are independently learned instead of multiplications between linear terms, is exhibited in \textbf{Fig. \ref{fig:fmqr}}.
These experiments are conducted with CRYSIM-5 representation of Y$_6$Co$_{51}$ system, containing 801 bits in the embeddings. Under these conditions, QR involves more than 600,000 trainable parameters, whereas FM requires only 13,617 ones.
For reference, optimization results of BO and CALYPSO, previously shown in \textbf{Table \ref{table:mainresults}} are also included as baselines, with three trials performed for each method.
On this system, CRYSIM-QR achieves a lower average accumulated energy than CRYSIM-FM, indicating superior optimization performance. 
%and successfully generated Y$_6$Co$_{51}$ structures close to the ground truth, as is shown in \textbf{Fig. \ref{fig:fmqr}b}, though energy of the one in the figure (-393.85 eV) is still higher than its stable state (-406.26 eV).
However, for systems represented with more than 1,000 bits, QR requires millions of trainable parameters, making FM a more practical option for these tasks in terms of computational efficiency.

\quad\\
\noindent\textbf{Processing with Minimum Interatomic Distance.}\quad
Inaccuracy of NNPs on configurations with extremely small MIDs renders negative impact on regression models.
To mitigate the effect, procedures related to MIDs are designed and integrated in the workflow, including selecting the structure with the largest MID from one solution vector in \textbf{Fig. {\ref{fig:workflow}}f}, and adjusting unphysical energy estimations in \textbf{Fig. {\ref{fig:workflow}}g}.
Effectiveness of the procedures is demonstrated in \textbf{Table \ref{table:as-pp}} by comparing CRYSIM of all considered LDRs with and without including these steps during optimization.
Tests of each method repeat three times with different seeds. The corresponding accumulated energy curves are provided in \textbf{Fig. \ref{fig:postuncons}}.
Optimizers equipped with the modules achieve a widespread performance enhancement, particularly for larger systems with higher LDRs.
Besides, inclusion of MID processing enables structures exploration with high LDRs. For Y$_6$Co$_{51}$ and Ca$_{24}$Al$_{16}$(SiO$_4$)$_{24}$, CRYSIM-12 performs the best with MID-related procedures, but it cannot realize the full potential and is outperformed by CRYSIM-9 without them.
The advantage is attributed to improved efficiency in obtaining valid crystals, indicated by a notable reduction of filtered-out configurations reported in \textbf{Table \ref{table:as-filternumpp}}.

\section{Discussion}

\noindent\textbf{Representing fractional coordinates by lattice splitting.}\quad
There are two main strategies for representing fractional coordinates as variables to be optimized. The first is to treat each coordinate ($x_i$, $y_i$ and $z_i$ for the $i$-th atom) as independent variables \cite{RN951, RN678}, and the second is to split the whole crystal lattice and use the derived discrete blocks in the 3D space to encode positions \cite{RN965, RN1141, RN988, RN1155}.
Most optimization methods based on Ising models adopt the second approach, as it aligns with the goal of achieving guaranteed optimal solutions through quantum annealing by fitting the system’s PES.
%By splitting the lattice and constructing an Ising model in which variables represent blocks in the real space, 1-order terms describe energy contributions of a single atom located at the place originated from external fields, and 2-order terms describe 2-body interactions between atoms, so that Ising models simulate the interatomic potential in a quadratic form.
By representing atomic positions via lattice splitting, an Ising model can encode physical interactions: first-order terms capture the energy contribution of a single atom due to external fields, while second-order terms describe pairwise atomic interactions. This allows the Ising model to approximate the interatomic potential in a quadratic form.
	
Nevertheless, in practical implementations that account for symmetry, such as CRYSIM and other works \cite{RN965}, the solved atomic positions are not external coordinates for building structures, but internal or independent sites to insert in WPs.
As a result, the learned Ising model does not fully reflect an actual interatomic potential.
One possible solution is to first estimate or sample an SG and WPC, derive corresponding constraints on lattice parameters and coordinates, and then optimize the two parts.
Accordingly, by adding penalty terms, Ising solver can optimize directly on external coordinates and preserving the symmetry simultaneously.
However, the order of constraints on coordinates would be the same as the multiplicity of corresponding WPs, making it challenging to implement for current solvers.

From a practical perspective, another advantage of the second strategy over the first one is that it requires less bits to encode coordinates in large systems, especially those with few atomic species.
This is because the second strategy scales linearly with the number of atomic species and remains constant with respect to the number of atoms, whereas the first strategy scales linearly with the number of atoms within the cell.
The scaling with the number of atoms is generally more computationally intensive in large systems.
Taking the (SiO$_2$)$_{96}$ system in \textbf{Results} as an example, suppose the lattice is split into $15*15*15$ blocks. Following the first strategy, each coordinates require 15 bits to represent, leading to $15 * 3 * (96 + 192) = 12960$ bits in total, but only $15 * 15 * 15 * 2 = 6750$ are needed based on the second one.

\quad\\
\noindent\textbf{MID-related procedures.}\quad Ideally, a pretrained NNP should assign high energies to unstable structures, allowing the objective functions to reflect an accurate structure-energy relationship through active learning for diversion structural configurations.
Accordingly, CSP optimizers, designed to identify low-energy solutions, can correctly discover the ground states.
However, training sets for state-of-the-art NNPs generally lack out-of-distribution structures in the configuration space, especially for crystals containing extremely close atom pairs, rendering their estimated energies unreasonably low. As presented in \textbf{Table \ref{table:as-filternumpp}}, many CRYSIM optimizers without MID processing are encouraged to generate abnormal structures, due to their low estimated relaxed energies, which are, however, ineffective from a practical perspective.
Directly replacing pretrained NNP with first-principles calculation software \cite{RN444, RN610} can circumvent the problem, but it is still unrealistic for large crystals considering current computational power.

According to an observation that most abnormal relaxed structures originate from abnormal decoded unrelaxed ones, we design MID processing techniques on generated configurations (\textbf{Fig. {\ref{fig:workflow}}f} and \textbf{g}) to improve efficiency of CRYSIM optimizers. Besides, previous works \cite{RN959} also suggest that atoms in stable structures tend to uniformly distribute, instead of clustering in a small space.
However, for some material systems, strategies aimed at controlling MIDs of generated materials (e.g., CRYSPY RG and CALYPSO) or attempting to obtain materials with larger MIDs (e.g., CRYSIM) may hinder discovering of ground states, as evidenced by the Y$_6$Co$_{51}$ system in \textbf{Table \ref{table:mainresults}}.
Although many structures derived by RG and BO implemented in this work are screened out due to very small MIDs, as reported in \textbf{Table \ref{table:filternumclassical}}, CSP of Y$_6$Co$_{51}$ is finally accomplished after structure relaxation on crystals that may not have high MIDs.
We expect that when a more accurate pretrained NNP is proposed, the MID-related procedures can be discarded.

\quad\\
\noindent\textbf{End-to-end CSP algorithm evaluation.}\quad
The primary objective of our optimization-based CSP algorithm, CRYSIM, is precisely to locate the global minimum on the PES, representing the most thermodynamically stable structure according to the chosen energy model.
To rigorously assess this specific capability during benchmarking, we consider the number of "successful match" ($N_M$) only for structures with the lowest relaxed energies out of the 300 total structures instead of all of them.
This prevents crediting success to fortuitous sampling into higher-energy local minima, thereby isolating the evaluation of optimization performance.
%The inherent complexity of PES for crystalline materials, particularly complex ones, results in a highly rugged landscape populated by numerous local minima.
%A slight vibration of atomic coordinates can lead to great change in energy.
%This complexity, coupled with potential inaccuracies in the underlying energy model, can lead to situations where the calculated energy ranking of predicted polymorphs does not perfectly align with structural similarity to the experimental ground truth.
%Consequently, there exists a possibility that a predicted structure residing in a higher-energy local minimum might, by chance, exhibit greater structural resemblance to the known ground truth than the structure identified as the global minimum candidate.
%When searching for successful predictions in all experiments, only the ones with the lowest relaxed energies out of the totally 300 structures are considered, and compared with their stable counterparts. Due to the high ruggedness of PES of complex materials, a slight vibration of atomic coordinates can lead to great change in energy. It is possible that a configuration with a higher estimated energy shares more similarity with the ground truth than another local minimum with lower energy. However, under the framework of CSP, those obtained configurations are not recognized as successful results, but attribute to randomness, since they do not reflect optimization power of the algorithm.

More critically, this methodology also reflects the practicability in CSP tasks.
When the target structure is unknown, researchers inevitably rely on the calculated energy ranking, treating the lowest-energy prediction as the most likely candidate for experimental synthesis or validation.
Higher-energy predictions, even if potentially correct, cannot be identified as such without prior knowledge of the ground truth. 
Thus, by focusing our evaluation on the lowest-energy structure, our metric is not only relevant to CRYSIM's optimization objective but also aligned with the practical interpretation and utility of CSP results.
%Besides, in real CSP tasks, stable states of a composition are unknown. Even if a generated material is similar to the actual ground truth, the fact cannot be unveiled, since the pair-wise displacement is uncomputable. The most stable generated crystal is the one most likely to be the globally stable state predicted by the model, and will be adopted practically. Therefore, only the configuration with the lowest energy should be considered when evaluating CSP algorithms.
%From this perspective, correct energy estimations are of great significance for the success of CSP.

\quad\\
\noindent\textbf{Leveraging quantum annealing.}\quad Quantum annealing (QA) \cite{RN1112, RN1113, RN1109, RN1102, RN1103, RN1104} has gained attention due to its theoretical ability to escape from local minima, and D-Wave system \cite{8728085} is among the most widely used implementations of QA \cite{RN965, RN988, RN1166}. However, the maximum number of variables for the D-Wave system is limited to 124 bits \cite{RN964}, severely restricting its application. In this work, we integrate Amplify \cite{RN1120}, a GPU-based Ising solver, into CRYSIM as a substitute of quantum annealers. Nevertheless, we claim that the present implementation can be applied directly on quantum annealers without adjustment as quantum computers continue to develop.

\quad\\
In conclusion, CRYSIM, a CSP optimizer based on symmetry-encoded Ising models, is proposed and tested across various CSP tasks.
To the best of our knowledge, it is the first Ising machine-based optimizer for CSP that dynamically optimizes symmetry.
CRYSIM outperforms CRYSPY RG, BO, and CALYPSO on most systems, showcasing its strong optimization capabilities not only for small benchmark crystals but also for larger ones, including Ca$_{24}$Al$_{16}$(SiO$_4$)$_{24}$ and (SiO$_2$)$_{96}$.
The predicting accuracy of FM in CRYSIM is also discussed, highlighting its expressivity in CSP tasks.
CRYSIM offers a promising Ising machine-based optimization tool for CSP that could potentially be applied to quantum annealers in the future.

\section{Methods}
\label{sec:Method}

\subsection{Random Generation of Crystal Structures}
\label{sec:rg}

Random generation (RG) constitutes the foundation of CSP \cite{RN1060, RN951, RN958}, in which atomic positions are randomly sampled according to the given chemical composition. In this work, a simple crystal RG tool is implemented as the CSP baseline, as well as preparing training data for Ising models.

In RG, six lattice parameters (lattice lengths $a$, $b$, $c$, and lattice angles $\alpha$, $\beta$, $\gamma$) and fractional coordinates are sampled independently from uniform distributions. 
To determine the default lower and upper bounds of the distributions for lattice lengths, we perform statistical analysis on materials in MP.
Let $M$ be the set of all materials in the MP database.
Define a function $\vert m\vert: M \to \mathbb{N}$ that maps each material $m$ to the number of atoms in its unit cell.
We then partition $M$ into five categories, $\{M_1, M_2, M_3, M_4, M_5\}$, such that for each material $m \in M$, it belongs to category $M_i$ if it satisfies
\begin{equation}
	\left\{
	\begin{aligned}
		M_1 &= \{ m \mid |m| \leq 20 \}, \\
		M_2 &= \{ m \mid 20 < |m| \leq 50 \}, \\
		M_3 &= \{ m \mid 50 < |m| \leq 80 \}, \\
		M_4 &= \{ m \mid 80 < |m| \leq 100 \}, \\
		M_5 &= \{ m \mid |m| > 100 \}.
	\end{aligned}
	\right.
\end{equation}
Next, the averages of $a$, $b$, and $c$ for materials in each category $M_i$ are computed, denoted as $a_{M_i}$, $b_{M_i}$, and $c_{M_i}$, respectively.
For a specific material system $m_0$ to be generated, if the number of atoms $\vert m_0\vert$ falls within one of the ranges, the lower and upper bounds are determined as follows:
\begin{equation}
	\left\{
	\begin{aligned}
		l_{m_0}&=0.8 \times (a_{M_i} + b_{M_i} + c_{M_i}),\\
		u_{m_0}&=2 \times (a_{M_i} + b_{M_i} + c_{M_i}).
	\end{aligned}
	\right.
\end{equation}
These bounds are the same for the three lattice lengths.
The lower and upper bounds for the lattice angles are set to $50^\circ$ and $130^\circ$, respectively.

Then, space group (SG) and corresponding Wyckoff positions combination (WPC) are derived for building symmetry.
Given the input chemical composition, let $\mathbb{S}_+$ be the set of SGs that are compatible with the stoichiometry. 
For each $S\in \mathbb{S}_+$, let $\mathbb{W}_S$ be the set of corresponding WPCs (see \textbf{Section \ref{sec:wpc}} of \textbf{Supplementary Information}).
Let $\vert\mathbb{W}_S\vert$ denote the number of distinct compatible WPCs for SG $S$.
The process involves sampling an SG and then a WPC.
An SG number is sampled from all compatible ones ($\mathbb{S}_+$) uniformly, i.e., an SG $S_l$ is selected by sampling its identifier $i_{S_l}$ uniformly from the set of identifiers for SGs in $\mathbb{S}_+$:
\begin{equation}
	i_{S_l} \sim \mathcal{U}(\{\text{id}(S) \mid S \in \mathbb{S}_+\}), \quad S_l = \text{SG}(i_{S_l}).
\end{equation}
Based on $S_l$, a WPC is subsequently sampled.
Define the maximum WPC count $W_{\max}=\max_{S\in \mathbb{S}_+} \vert\mathbb{W}_S\vert$.
Sample an integer $i_{W_l}$ uniformly from $\{0, 1, \dots, W_{\max}-1\}$:
\begin{equation}
	i_{W_l} \sim \mathcal{U}(\{0, 1, \dots, W_{\max}-1\}).
\end{equation}
The WPC $W_l$ is derived from the WPCs set corresponding to the chosen SG $S_l$ based on the sampled identifier:
\begin{equation}
	W_l = \text{WPC}_{S_l}(\left\lfloor i_{W_l} \cdot \frac{\vert\mathbb{W}_{S_l}\vert}{W_{\max}} \right\rfloor),
\end{equation}
where $\lfloor x \rfloor$ is the floor function, which returns the greatest integer less than or equal to $x$.

Finally, fractional coordinates are uniformly sampled from the interval $[0, 1)$.
These coordinates are treated as independent sites and placed into the Wyckoff positions $W_l$, where they are transformed into external coordinates to satisfy symmetry constraints.
Similarly, the generated lattice parameters are assigned to variables defined by the crystal system (CS) $C_l$ associated with the sampled SG $S_l$.
Since WPCs impose dependencies among coordinates, reducing the degrees of freedom, only the earliest generated coordinates are used.
The same approach applies to lattice parameters constrained by a CS.

%Generated crystals should be assigned with symmetry in structures, i.e., accordant with some SG, to increase the possibility of discovering stable states. 
%Given an SG and number of atoms, a set of corresponding Wyckoff positions (WPs) combinations can be inferred, which define all possible coordinates that atoms in unit cells can occupy to fulfill symmetry of the SG.
%Taking the A$_4$B$_4$ system as an example, if four A and B atoms in a configuration satisfying the following relationship denoted by a set of WPs combination: 
%\begin{equation}
%	\left\{
%	\begin{aligned}
%		A_1&: (x_1, y_1, z_1),\\
%		A_2&: (-x_1, -y_1, z_1+1/2),\\
%		A_3&: (-y_1, x_1, z_1+3/4),\\
%		A_4&: (y_1, -x_1, z_1+1/4),\\
%		B_1&: (x_2, y_2, z_2),\\
%		B_2&: (-x_2, -y_2, z_2+1/2),\\
%		B_3&: (-y_2, x_2, z_2+3/4),\\
%		B_4&: (y_2, -x_2, z_2+1/4),
%	\end{aligned}
%	\right.
%\end{equation}
%the structure has symmetry defined by the $P4_3$ SG.
%In our implementation, based on the number of atoms and stoichiometry of the current system, WPs combinations for totally 229 SGs are calculated and saved before generation. When randomly generating a structure, lattice parameters, fractional coordinates, SG ($S_0$), simultaneously the crystal system (CS, $C_0$), and a WPs combination ($W_0$) are first randomly sampled. 
%Then, the lattice parameters and fractional coordinates are filled into the variables indicated by $C_0$ and $W_0$, so that the output structure has symmetry with respect to $S_0$. The implementation is based on GN-OA \cite{RN678}. 

When generating datasets for training, structures containing atom pairs with distances smaller than 1.5 \AA\ are removed to ensure that most generated structures have a reasonable estimated energy, which is essential for training an accurate objective function. Distance filtering is not involved when evaluating performance of the RG baseline. A much refined RG process is implemented by PyXtal \cite{RN957}, which has been tested as the CRYSPY RG baseline in this work.

\subsection{Details of Symmetry-informed Integer Encoding in CRYSIM}

Integer encoding can be interpreted as follows. Suppose a binary vector segment containing $N_{v}$ bits is leveraged for representing parameter $v$. For a continuous parameter $v \in [v_{\mathrm{min}}, v_{\mathrm{max}})$, the $i_v$-th bit of the vector segment will be assigned as 1 and other elements are 0s with
\begin{align}
	i_v = \left\lfloor \frac{v - v_{\text{min}}}{u_v} \right\rfloor,
	\label{eq:onehot}
\end{align}
where $u_v = (v_{\mathrm{max}} - v_{\mathrm{min}}) / N_v$ is the unit or interval of the representation. For a discrete parameter, $i_v=1$ if the parameter $v$ of the system corresponds to the $i$-th category.

\quad\\
\noindent\textbf{Lattice parameters encoding.}\quad In the workflow of CRYSIM, integer encoding is initially performed on training sets generated by RG.
The upper and lower bound of lattice length encoding, simultaneously the highest and lowest lattice length of decoded materials, are calculated based on data points in the sets.
Let $ll_{0,\max}$ and $ll_{0,\min}$ represent the maximum and minimum lattice lengths (for $a$, $b$, and $c$) among all structures in the initial training set.
Then, the upper and lower bounds for lattice length encoding are defined as
\begin{equation}
	\left\{
	\begin{aligned}
		ll_{\mathrm{max}}&= 1.1 * ll_{0, \mathrm{max}},\\
		ll_{\mathrm{min}}&= ll_{0, \mathrm{min}}.
	\end{aligned}
	\right.
\end{equation}
The number of bits for representing lattice lengths $N_a=N_b=N_c=N_{ll}$ is dependent on divisions of the lattice when encoding atomic coordinates, which is calculated by
\begin{equation}
	N_{ll} = C_{ll} * g * \frac{ll_{\mathrm{max}} - ll_{\mathrm{min}}}{ll_{\mathrm{max}}},
\end{equation}
in which $g$ denotes the current LDR, and $C_{ll} = 10$ by default.
For lattice angles ($\alpha$, $\beta$ and $\gamma$), the lower and upper bound are $50^\circ$ and $130^\circ$, the same as RG. The unit for encoding angles is $2^\circ$, so that one lattice angle is encoded using
\begin{equation}
	N_{lg} = (130 - 50) / 2 = 40
\end{equation}
bits. The number of bits for encoding lattice parameters would be $3 * N_{ll} + 3 * N_{lg}$.

\quad\\
\noindent\textbf{Fractional coordinates encoding.}\quad Atomic configurations are represented by discretizing the unit cell into a $g \times g \times g$ voxel grid, where $g$ is the LDR. A 3D binary matrix $X \in \{0, 1\}^{g \times g \times g}$ is constructed, where element $X_{l,m,n}$ corresponds to the voxel region $R_{l,m,n}$ defined by fractional coordinates $\mathbf{y} = (y_1, y_2, y_3)$:
\begin{equation}
	R_{l,m,n} = \left\{ \mathbf{y} \mid \frac{l}{g} < y_1 < \frac{l+1}{g}, \; \frac{m}{g} < y_2 < \frac{m+1}{g}, \; \frac{n}{g} < y_3 < \frac{n+1}{g} \right\}
\end{equation}
for $l, m, n \in \{0, 1, \dots, g-1\}$. $X_{l,m,n}$ is set to 1 if an atom's fractional coordinates fall within $R_{l,m,n}$, and 0 otherwise.
For crystals containing multiple element species, a separate flattened matrix is constructed for each of them.
After concatenation, encoded information of each element is stored in separate regions of the final embedding.

We note that the optimized coordinates are internal ones, which will be placed into WPCs to satisfy symmetry constraints.
Besides, similar to RG, the derived bits might be redundant. In implementation, 1-bits in the leftmost positions in the vector segment for each element are used, until all variables in the solved WPC are decided.
For experiments on benchmark crystals, the Y$_{6}$Co$_{51}$ and Ca$_{24}$Al$_{16}$(SiO$_4$)$_{24}$ system in this study, LDRs of CRYSIM are set to 12, with (SiO$_2$)$_{96}$ being 15.

\quad\\
\noindent\textbf{Symmetry information representation.}\quad
Symmetry information involves the CS, SG and WPC, which define the crystal’s symmetry. Numbers of bits for the three parts, i.e., $N_{C}$, $N_{S}$ and $N_{W}$, are dependent on the WPCs list calculated based on stoichiometry of the system.
To be specific, only compatible SGs and CSs are encoded, and whether an SG and CS is compatible or not is determined by the existence of WPCs that can be used to build the corresponding symmetry, as is illustrated in \textbf{Section \ref{sec:wpc}} of \textbf{Supplementary Information}.
% For instance, there is no possible WPs combinations available for any A$_4$B$_4$ systems given the $F4_132$ SG, thus this type of crystals can never have the symmetry. Accordingly, the $F4_132$ SG should not be included in the embedding segment. Furthermore, availability of CS relies on SGs. In A$_3$B$_3$C$_3$ systems, none of SG numbers within $[195, 230]$ can be achieved, so that this materials family cannot have cubic lattices.

Details of calculating $N_{C}$, $N_{S}$ and $N_{W}$ are presented as follows. 
%First of all, WPs combinations for all 229 SGs are calculated, denoted as $\mathbb{W}_{all}=\{\mathbb{W}_{S}|S\in \mathbb{S}_{all}\}$, in which $\mathbb{S}_{all}$ includes all SGs from 2 to 230.
Let $\mathbb{C}_+$ and $\mathbb{S}_+$ denote all compatible CSs and SGs, respectively, and $\mathbb{S}_{C}$ denotes the set of SGs associated with a CS $C$.
We then define the set of compatible SGs for $C$ as $\mathbb{S}_{C, +} = \mathbb{S}_{C} \cap \mathbb{S}_{+}$.
The numbers of bits for representing CSs and SGs can be decided as
\begin{equation}
	\left\{
	\begin{aligned}
		N_{C}&=|\mathbb{C}_+|,\\
		N_{S}&=\max_{C \in \mathbb{C}_+} |\mathbb{S}_{C,+}|.
	\end{aligned}
	\right.
\end{equation}
$N_{W}$ can be independently set.
In many cases, the number of distinct compatible WPCs for an SG $S$, denoted as $\vert\mathbb{W}_{S}\vert$, is so large that encoding all WPCs within a binary segment becomes impractical.
To address this, only WPCs with indices within the set $\left\{ \left\lfloor \frac{i \cdot \vert\mathbb{W}_{S}\vert}{N_W} \right\rfloor \mid i = 0, 1, \dots, N_W - 1 \right\}$ are included.
In this work, $N_W$ is set to 300 by default.
However, during decoding, each bit $i$ corresponds to a group of 100 WPCs with indices $\left\{ \left\lfloor \frac{i \cdot \vert\mathbb{W}_{S}\vert}{N_W} \right\rfloor + j \mid j = 0, 1, \dots, 99 \right\}$.

When encoding symmetry information of a crystal having CS $C_l$, SG $S_l$ and WPC $W_l$, first of all, the corresponding bit representing the CS is assigned as 1.
The bit for $S_l$ is derived by
\begin{equation}
	i_{S_l} = \lfloor \text{id}_{C_l,+}(S_l) * \frac{N_S}{|\mathbb{S}_{C_l,+}|} \rfloor,
	\label{eq:onehotsg}
\end{equation}
in which $\text{id}_{C_l,+}(S_l)$ is the identifier of $S_l$ in the ascending ordered sequence with respect to the set $\mathbb{S}_{C_l,+}$. For instance, in the SG set of the cubic CS, i.e., $\{P23, F23, ..., Ia\overline{3}d\}$, the identifier of SG $P23$ is 0.
Similarly, the bit for $W_l$ is calculated by
\begin{equation}
	i_{W_l} = \lfloor \text{id}_{S_l}(W_l) * \frac{N_{W}}{|\mathbb{W}_{S_l}|} \rfloor.
	\label{eq:onehotwy}
\end{equation}
where $\text{id}_{S_l}(W_l)$ represents the identifier of $W_l$ in $\mathbb{W}_{S_l}$.
%, but the value has been magnified to the same level of $W_{\mathrm{max}}$ by multiplying $W_{\mathrm{max}} / |\mathbb{W}_{S}|$. For instance, if $W_{\mathrm{max}}=1,000$, and $|\mathbb{W}_{S}|=2$, then $W_{idx,S}$ for the two WPs combinations in the set would be $0$ and $500$.
%In some cases, $|\mathbb{W}_{S}|$ is approaching one million, so that most of WPs combinations are not included in the $\mathcal{N}_{W}$ bits. However, all of the WPs in $\mathbb{W}_{S}$ are equivalent since they lead to the same symmetry defined by $S$.
%The WPs not encoded are further utilized in the post-processing steps.

During decoding processes, the following relationships are adopted
\begin{equation}
	\left\{
	\begin{aligned}
		\text{id}_{C_l,+}(S_l) =& \lfloor i_{S_l} * \frac{|\mathbb{S}_{C,+}|}{N_{S}} + 0.5 \rfloor, \quad S_l = \text{SG}_{C_l,+}(\text{id}_{C_l,+}(S_l)), \\
		\text{id}_{S_l}(W_l) =& \lfloor i_{W_l} * \frac{|\mathbb{W}_{S}|}{N_{W}} + 0.5 \rfloor, \quad W_l = \text{WPC}_{S_l}(\text{id}_{S_l}(W_l)),
	\end{aligned}
	\right.
	\label{eq:onehotde}
\end{equation}
to make the encoding-decoding procedures stable and reversible.

We note that according to the integer encoding strategy implemented in CRYSIM, only one bit should be assigned as 1 in a vector segment for one parameter, so that decoding from the segment to real values can be performed directly. Amplify provides options to add constraints as penalty terms to the objective function to encourage generation of solutions fulfilling specific requirements.
In summary, the number of bits for symmetry encoding would be $N_{C} + N_{S} + N_{W}$.

\quad\\
\noindent\textbf{Example on symmetry information representation.}\quad
We further provide an illustrative example on encoding and decoding symmetry information. For the A$_4$B$_4$ system, there are 2 SGs available for the triclinic CS, 12 for monoclinic, 56 for orthorhombic, 66 for tetragonal, 12 for trigonal, 19 for hexagonal, and 17 for cubic. Since all CSs are compatible with the system, $N_{C}=7$. The maximum number of compatible SGs across all CSs is 66, and therefore $N_{S}=66$. $ N_{W}$ can be independently set as 300 by default. Accordingly, the total number of bits for encoding symmetry information is 7 + 66 + 300.
When encoding SGs, as an example, $P4_1$, the No.76 SG, is the second compatible SG belonging to tetragonal CS, thus the $4_{th}$ bit for CS and the $2_{nd}$ bit for SG are set to 1. For $P23$ (SG No.195) and $F23$ (SG No. 196), the first and second compatible SG in the cubic category, the corresponding bit for SG is calculated following
\begin{equation}
	i_{P23} = \lfloor 0 * \frac{66}{17} \rfloor = 0,
\end{equation}
and 
\begin{equation}
	i_{F23} = \lfloor 1 * \frac{66}{17} \rfloor = 3,
\end{equation}
respectively. 
On the other hand, in the decoding process, the bit 2, 3 and 4 will correspond to
\begin{equation}
	\left\{
	\begin{aligned}
	\lfloor 2 * \frac{17}{66} + 0.5 \rfloor = 0 \rightarrow P23, \\
	\lfloor 3 * \frac{17}{66} + 0.5 \rfloor = 1 \rightarrow F23, \\
	\lfloor 4 * \frac{17}{66} + 0.5 \rfloor = 1 \rightarrow F23.
	\end{aligned}
	\right.
\end{equation}

In actual implementation, since it is impossible to determine WPC index from an already generated structure, we only curate training sets obtained from our RG algorithm, in which crystals are constructed based on an already chosen WPC.

% spglib \cite{RN972} is employed for deciding 

\subsection{Criterion for Matching Structures}

\texttt{StructureMatcher} function in \texttt{pymatgen} package \cite{ONG2013314} is used to compare configurations, in which parameters are set as \texttt{stol=0.5}, \texttt{ltol=0.3}, \texttt{angle\_tol=10.0}, consistent with other related works \cite{jiao2023crystal, sriram2024flowllm}.
This function will calculate the minimum normalized average root mean square pair-wise displacement between two input structures among all atom permutations. But if corresponding atoms in the two structures are not detected, which means that the function cannot identify any similarity between them, the calculation will not be proceeded. Accordingly, a structure is recognized to be accordant with the ground truth if having a computable displacement with it, and a model successfully finds the ground truth in one run if there is at least one such structure being generated.

\subsection{Factorization Machine for Quadratic Regression}
\label{sec:fm}

Factorization machine (FM) is a type of regression model proposed as a substitute of Support Vector Machine to address its failure on sparse data \cite{5694074}.
%The method has been successfully applied in materials and chemistry topics including optimizing nanowire structures for specific spectral absorbing preference \cite{RN1110}, organic molecules for various properties \cite{RN964} and high entropy alloys structure prediction \cite{RN1155}.
FM creates a mapping between a vector $\mathbf{x}\in \mathbb{R}^{M}$ and real value $y$ by
\begin{equation}
	y = b + \sum_{i=1}^M h_i x_i + \sum_{i,j=1}^M
	\sum_{k=1}^K w_{ki} w_{kj} x_i x_j,
\end{equation}
%\begin{align}
%	y = w_0 + \sum_{i=1}^n w_{1, i} x_i + \sum_{i=1}^n \sum_{j=i+1}^n \langle \mathbf{w}_{2, i}, \mathbf{w}_{2, j} \rangle x_i x_j,
%	\label{eq:fm}
%\end{align}
where $b$, $h_{i}$, ${w}_{k, i}$ are coefficients for bias, linear and quadratic interactions, respectively. In principle, FM can be extended to model $n$-interaction terms, but we restrict it to quadratic terms since current combinatorial optimizers are efficient only for solving quadratic objective functions. In that case, FM can be reformulated as
\begin{align}
	y = b + \sum_{i=1}^M h_{i} x_i + \frac{1}{2} \sum_{k=1}^K \left( \left( \sum_{i=1}^M w_{ki} x_i \right)^2 - \sum_{i=1}^M w_{ki}^2 x_i^2 \right),
	\label{eq:fm2}
\end{align}
reducing computational complexity from $O(KM)$ to $O(2K)$ \cite{5694074}. One of advantages of FM in this work is that it requires less fitting parameters, enabling a quadratic regression on binary vectors containing thousands of bits. Taking a vector of 2,000 bits as an example, a full-rank quadratic regressor requires $2000\times 1999$ terms for interactions, while FM only needs $2000\times K$ terms, in which $K$ is usually smaller than 30. In this work, we implement FM with PyTorch \cite{NEURIPS2019_bdbca288} based on equation \ref{eq:fm2} to the accelerate learning process. 

\section*{Data availability}

Ground state configurations considered in this study can be downloaded from the MP database \cite{RN743}.
Initial datasets for training FM are generated using RG implemented in CRYSIM, and no external data is included.

\section*{Code availability}

Implementation of CRYSIM is available at \url{https://github.com/tsudalab/CRYSIM}. As of March 2025, Fixstars Amplify is available via Python API free of charge.

\section*{Acknowledgements}

K.T. is supported by JST ERATO JPMJER1903 and JST CREST JPMJCR21O2.
D.D. is supported by JSPS KAKENHI Young Scientist (23K16942).
C.L. would like to gratefully acknowledge the financial support from the China Scholarship Council (CSC No. 202306210120).
The authors thank Yaotang Zhang for discussions.

\section*{Author contributions statement}

C.L. implemented the CRYSIM package and conducted all experiments. D.D. contributed to insights into the design of Ising models and constraints. Z.M. contributed to implementation of the training framework of FM. J.G. and R.T. contributed to analysis about the usage of Amplify and other Ising solvers. K.T. proposed the idea of the work. Z.M. and K.T. provided guidance on experiments design and results analysis. All authors reviewed and contributed to the writing of the manuscript.

\section*{Additional information}

\textbf{Competing interests}: the authors declare no conflict of interest. 

%Bibliography
\phantomsection
\addcontentsline{toc}{section}{Reference}
\bibliographystyle{unsrt}  
\bibliography{references}

\clearpage

\appendix
\section*{Supplementary Information}

% Redefine figure numbering
\renewcommand{\thefigure}{S\arabic{figure}}
\setcounter{figure}{0}
\renewcommand{\thetable}{S\arabic{table}}
\setcounter{table}{0}

\tableofcontents

\clearpage

	\section{Supplementary Figures}
	
	\begin{figure}[htbp]
		\vspace{2.5em}
		\centering
		\includegraphics[width=\linewidth, center]{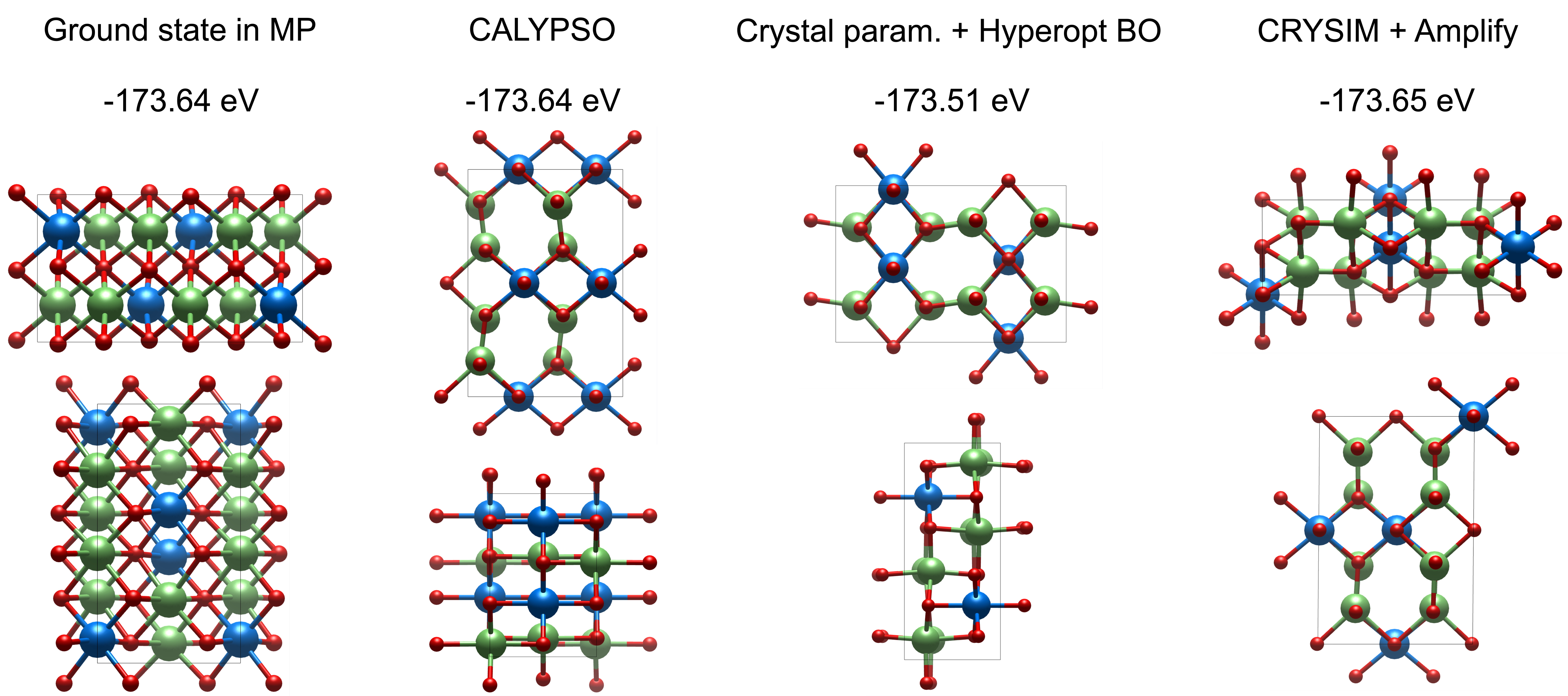}
		\caption{Side view (up row for each method) and top view (down row) of the ground state of Li$_8$Zr$_4$O$_{12}$ in MP (mp-4156, Li in green, Zr in blue, O in red), and predicted configurations by three CSP methods after structure relaxation, visualized by VESTA software \cite{Momma:db5098}, with M3GNet \cite{RN1130}-estimated relaxed energies labeled above.}
		\label{fig:li8zr4o12}
	\end{figure}
	
	\begin{figure}[htbp]
		\centering
		\includegraphics[width=\linewidth, center]{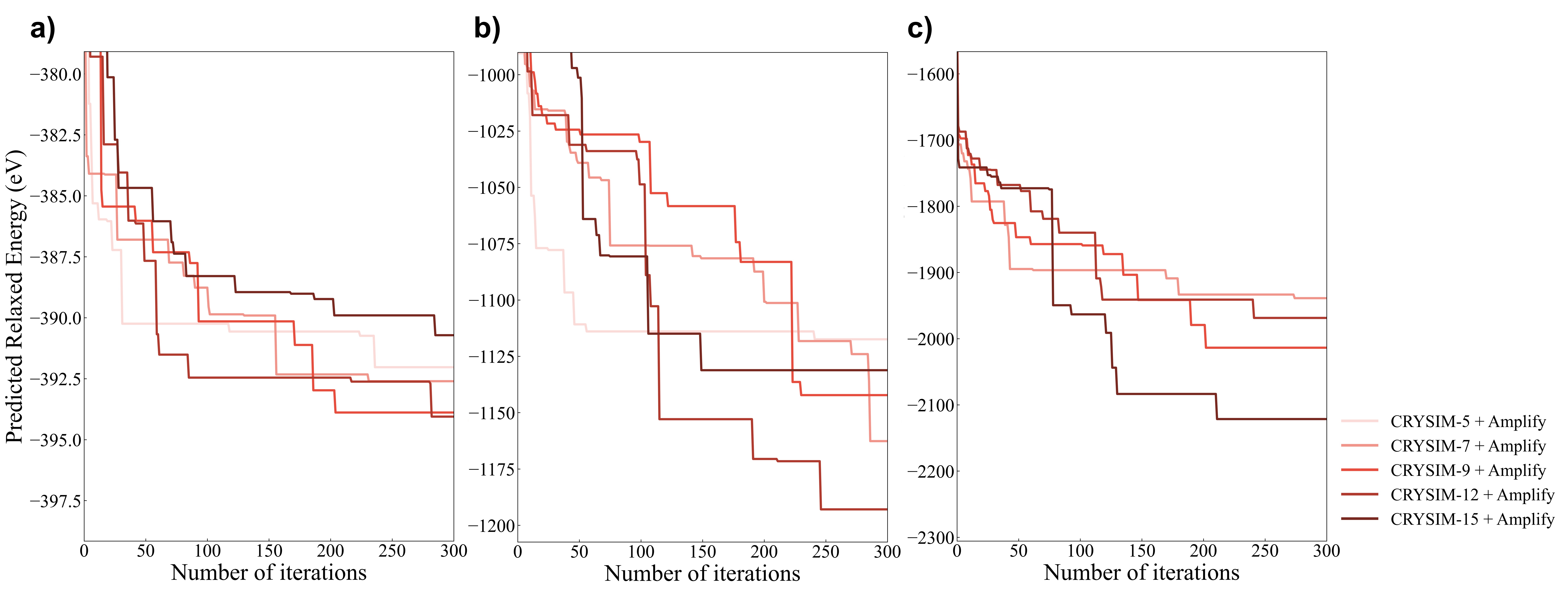}
		\caption{Averaged accumulated lowest M3GNet-estimated relaxed energies of structures generated various CRYSIM optimizers on \textbf{a} Y$_6$Co$_{51}$,  \textbf{b} Ca$_{24}$Al$_{16}$(SiO$_4$)$_{24}$, and \textbf{c} (SiO$_2$)$_{96}$ system, respectively. The number after "CRYSIM" in the legend indicate different LDRs. A deeper color suggests a higher LDR. Each curve is averaged on three tests with different random seeds.}
		\label{fig:splits}
	\end{figure}
	
	\clearpage
	
	\begin{figure}[htbp]
		\centering
		\includegraphics[width=\linewidth, center]{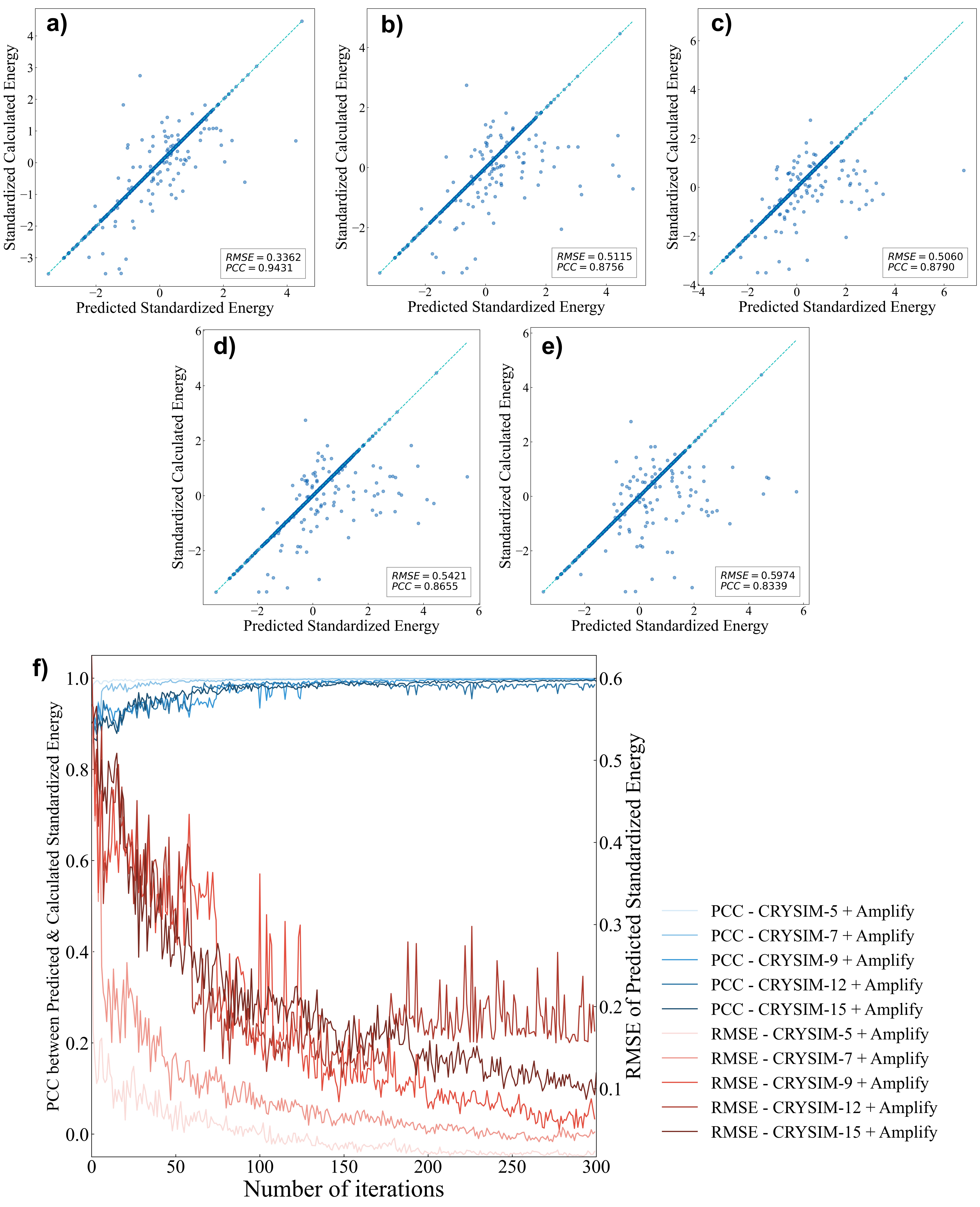}
		\caption{Performance on predicting energies of Ca$_{24}$Al$_{16}$(SiO$_4$)$_{24}$ dataset with the seed 0 using FM as the regressor. M3GNet-estimated energy versus predicted energy on the initial dataset, using \textbf{a} CRYSIM-5, \textbf{b} CRYSIM-7, \textbf{c} CRYSIM-9, \textbf{d} CRYSIM-12 and \textbf{e} CRYSIM-15 to encode crystal structures, in which Pearson correlation coefficients (PCCs) between calculated and predicted energies and root mean square errors (RMSEs) of predicted results are depicted in blue and red colors, respectively. \textbf{f} Tendency of PCCs and RMSEs during training as more and more structures included into the training set during active learning. Before used for learning, energies of crystals have been standardized to facilitate predicting accuracy.}
		\label{fig:fm}
	\end{figure}
	
	\clearpage
	
	\begin{figure}[htbp]
		\centering
		\includegraphics[width=\linewidth, center]{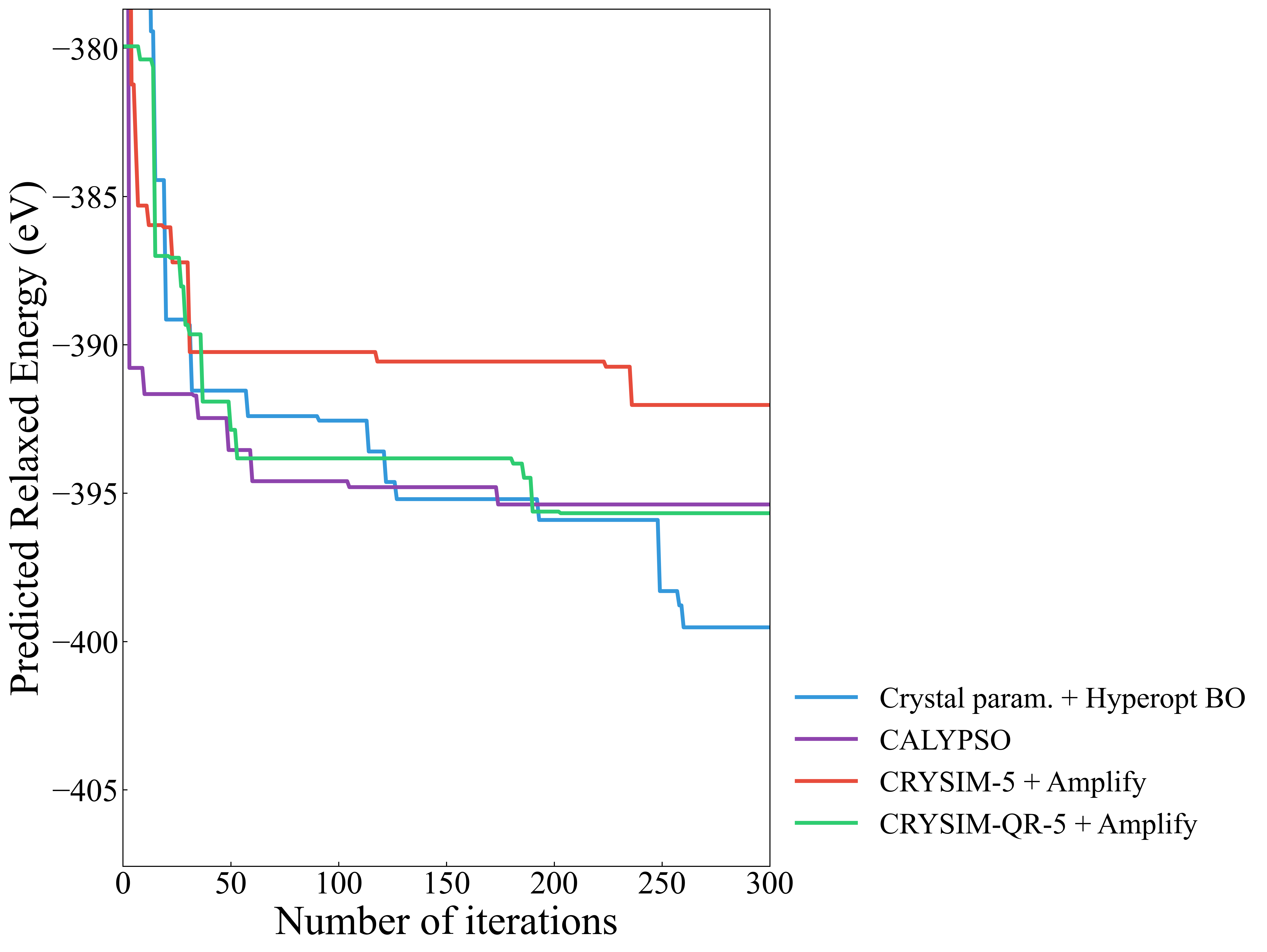}
		\caption{Comparison between CRYSIM-5 optimizers using Factorization Machine (FM) and full-rank quadratic regression (QR) as regressors on Y$_6$Co$_{51}$ system, denoted as "CRYSIM-5" and "CRYSIM-QR-5", respectively. Accumulated energy curves of BO and CALYPSO on the material are also included as baselines. Each curve is averaged on three tests with different random seeds.}
		\label{fig:fmqr}
	\end{figure}
	
	\clearpage
	
			\begin{figure}[htbp]
		%\vspace{4em}
		\centering
		\includegraphics[width=\linewidth, center]{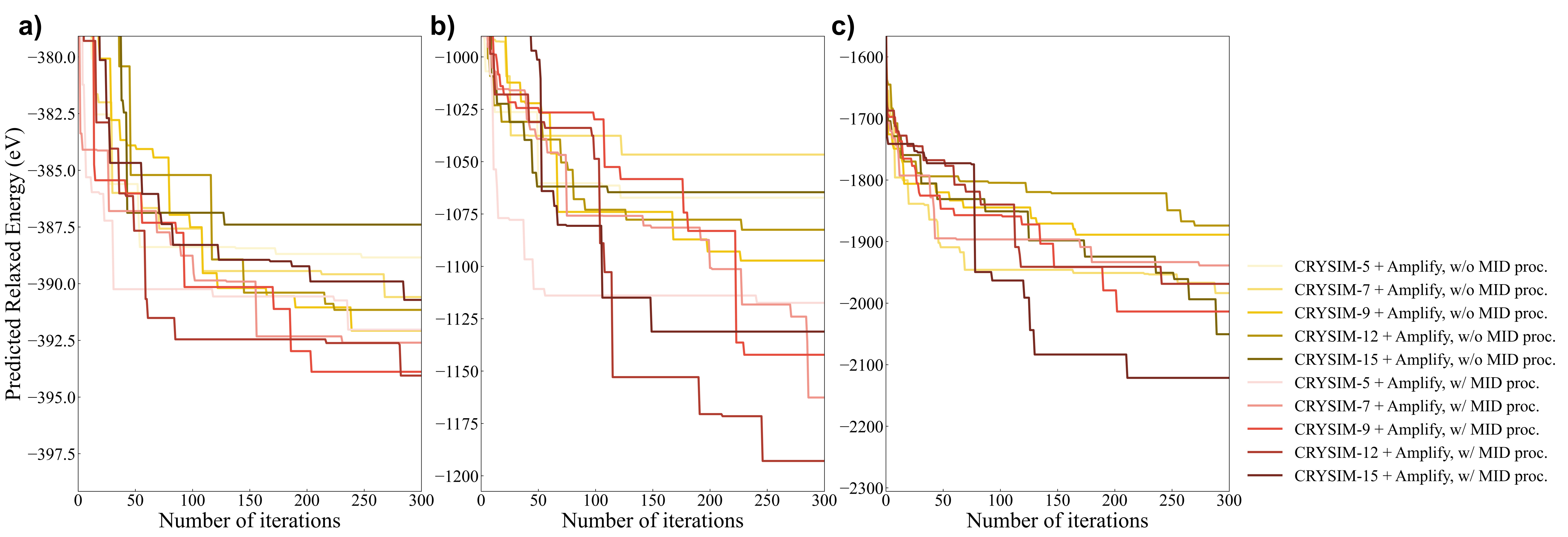}
		\caption{Comparison between averaged accumulated lowest M3GNet-estimated relaxed energies of structures generated by CRYSIM optimizers with and without integrating MID-related procedures on \textbf{a} Y$_6$Co$_{51}$,  \textbf{b} Ca$_{24}$Al$_{16}$(SiO$_4$)$_{24}$, and \textbf{c} (SiO$_2$)$_{96}$ system, respectively. Methods in which MID procedures are included, denoted as "w/ MID proc.", are drawn in red colors, otherwise, denoted as "w/o MID proc.", in yellow. Each curve is averaged on three tests with different random seeds.}
		\label{fig:postuncons}
	\end{figure}
	
	\clearpage
	
	\section{Supplementary Tables}
	
		\begin{table}[h]
		\caption{Comparison on ScBe$_5$ benchmark crystal, whose estimated stable energy is -26.24 eV.}
		\renewcommand\arraystretch{1.5}
		\centering
		\resizebox{14cm}{!}{
			\begin{tabular}{ccccc}
\hline
\multirow{2}{*}{Seed} & \multirow{2}{*}{Metrics} & \multicolumn{3}{c}{Optimizer} \\
~ & ~ & Crystal param. + Hyperopt BO & CALYPSO & CRYSIM + Amplify \\
\hline
\multirow{9}{*}{0} & 
$I_{M, 0}$ & 16 & 7 & 5 \\
~ & $D_{M, 0}$ & 0.00041 & 0.00058 & 0.00067 \\
~ & $I_{M, \min}$ & 277 & 8 & 187 \\
~ & $D_{M, \min}$ & 0.0 & 3e-05 & 0.00027 \\
~ & $E_{\min}$ (eV) & -26.24 & -26.24 & -26.24 \\
~ & $N_E$ & 12 & 62 & 170 \\
~ & $N_M$ & 12 & 62 & 170 \\
\hline
\multirow{9}{*}{1} & 
$I_{M, 0}$ & 42 & 7 & 2 \\
~ & $D_{M, 0}$ & 0.00076 & 0.00058 & 0.00698 \\
~ & $I_{M, \min}$ & 68 & 8 & 35 \\
~ & $D_{M, \min}$ & 0.0 & 3e-05 & 0.00031 \\
~ & $E_{\min}$ (eV) & -26.24 & -26.24 & -26.24 \\
~ & $N_E$ & 10 & 63 & 86 \\
~ & $N_M$ & 10 & 63 & 86 \\
\hline
\multirow{9}{*}{2} & 
$I_{M, 0}$ & 210 & 7 & 12 \\
~ & $D_{M, 0}$ & 0.01573 & 0.00058 & 0.00179 \\
~ & $I_{M, \min}$ & 254 & 8 & 187 \\
~ & $D_{M, \min}$ & 0.0 & 3e-05 & 0.00018 \\
~ & $E_{\min}$ (eV) & -26.24 & -26.24 & -26.24 \\
~ & $N_E$ & 5 & 54 & 60 \\
~ & $N_M$ & 5 & 54 & 60 \\
\hline
			\end{tabular}
		}
		\label{table:bench-scbe5}
	\end{table}
	
	\clearpage

	\begin{table}[h]
		\caption{Comparison on Ca$_4$S$_4$ benchmark crystal, whose estimated stable energy is -41.59 eV.}
		\renewcommand\arraystretch{1.5}
		\centering
		\resizebox{14cm}{!}{
			\begin{tabular}{ccccc}
\hline
\multirow{2}{*}{Seed} & \multirow{2}{*}{Metrics} & \multicolumn{3}{c}{Optimizer} \\
~ & ~ & Crystal param. + Hyperopt BO & CALYPSO & CRYSIM + Amplify \\
\hline
\multirow{9}{*}{0} & 
$I_{M, 0}$ & 24 & 8 & 4 \\
~ & $D_{M, 0}$ & 0.0 & 0.0 & 0.0 \\
~ & $I_{M, \min}$ & 24 & 8 & 4 \\
~ & $D_{M, \min}$ & 0.0 & 0.0 & 0.0 \\
~ & $E_{\min}$ (eV) & -41.59 & -41.59 & -41.59 \\
~ & $N_E$ & 21 & 23 & 199 \\
~ & $N_M$ & 21 & 23 & 199 \\
\hline
\multirow{9}{*}{1} & 
$I_{M, 0}$ & 29 & 1 & 2 \\
~ & $D_{M, 0}$ & 0.0 & 0.0 & 0.0 \\
~ & $I_{M, \min}$ & 29 & 1 & 2 \\
~ & $D_{M, \min}$ & 0.0 & 0.0 & 0.0 \\
~ & $E_{\min}$ (eV) & -41.59 & -41.59 & -41.59 \\
~ & $N_E$ & 34 & 35 & 170 \\
~ & $N_M$ & 34 & 35 & 170 \\
\hline
\multirow{9}{*}{2} & 
$I_{M, 0}$ & 11 & 1 & 1 \\
~ & $D_{M, 0}$ & 0.00075 & 0.0 & 0.0 \\
~ & $I_{M, \min}$ & 17 & 1 & 1 \\
~ & $D_{M, \min}$ & 0.0 & 0.0 & 0.0 \\
~ & $E_{\min}$ (eV) & -41.59 & -41.59 & -41.59 \\
~ & $N_E$ & 40 & 21 & 170 \\
~ & $N_M$ & 40 & 21 & 170 \\
\hline
			\end{tabular}
		}
		\label{table:bench-ca4s4}
	\end{table}
	
	\clearpage
	
	\begin{table}[h]
		\caption{Comparison on Ba$_3$Na$_3$Bi$_3$ benchmark crystal, whose estimated stable energy is -27.96 eV.}
		\renewcommand\arraystretch{1.5}
		\centering
		\resizebox{14cm}{!}{
			\begin{tabular}{ccccc}
\hline
\multirow{2}{*}{Seed} & \multirow{2}{*}{Metrics} & \multicolumn{3}{c}{Optimizer} \\
~ & ~ & Crystal param. + Hyperopt BO & CALYPSO & CRYSIM + Amplify \\
\hline
\multirow{9}{*}{0} & 
$I_{M, 0}$ & / & / & 153 \\
~ & $D_{M, 0}$ & / & / & 0.01669 \\
~ & $I_{M, \min}$ & / & / & 153 \\
~ & $D_{M, \min}$ & / & / & 0.01669 \\
~ & $E_{\min}$ (eV) & -27.81 & -27.81 & -27.96 \\
~ & $N_E$ & 12 & 1 & 1 \\
~ & $N_M$ & 0 & 0 & 1 \\
\hline
\multirow{9}{*}{1} & 
$I_{M, 0}$ & / & / & 156 \\
~ & $D_{M, 0}$ & / & / & 0.00671 \\
~ & $I_{M, \min}$ & / & / & 156 \\
~ & $D_{M, \min}$ & / & / & 0.00671 \\
~ & $E_{\min}$ (eV) & -27.81 & -27.81 & -27.96 \\
~ & $N_E$ & 3 & 1 & 3 \\
~ & $N_M$ & 0 & 0 & 3 \\
\hline
\multirow{9}{*}{2} & 
$I_{M, 0}$ & / & / & / \\
~ & $D_{M, 0}$ & / & / & / \\
~ & $I_{M, \min}$ & / & / & / \\
~ & $D_{M, \min}$ & / & / & / \\
~ & $E_{\min}$ (eV) & -27.8 & -27.81 & -27.32 \\
~ & $N_E$ & 1 & 1 & 1 \\
~ & $N_M$ & 0 & 0 & 0 \\
\hline
			\end{tabular}
		}
		\label{table:bench-banabi}
	\end{table}
	
	\clearpage
	
	\begin{table}[h]
		\caption{Comparison on Li$_4$Zr$_4$O$_8$ benchmark crystal, whose estimated stable energy is -123.16 eV.}
		\renewcommand\arraystretch{1.5}
		\centering
		\resizebox{14cm}{!}{
			\begin{tabular}{ccccc}
\hline
\multirow{2}{*}{Seed} & \multirow{2}{*}{Metrics} & \multicolumn{3}{c}{Optimizer} \\
~ & ~ & Crystal param. + Hyperopt BO & CALYPSO & CRYSIM + Amplify \\
\hline
\multirow{9}{*}{0} & 
$I_{M, 0}$ & 272 & / & / \\
~ & $D_{M, 0}$ & 0.0124 & / & / \\
~ & $I_{M, \min}$ & 281 & / & / \\
~ & $D_{M, \min}$ & 0.0117 & / & / \\
~ & $E_{\min}$ (eV) & -123.17 & -123.4 & -123.4 \\
~ & $N_E$ & 3 & 1 & 2 \\
~ & $N_M$ & 3 & 0 & 0 \\
\hline
\multirow{9}{*}{1} & 
$I_{M, 0}$ & / & / & / \\
~ & $D_{M, 0}$ & / & / & / \\
~ & $I_{M, \min}$ & / & / & / \\
~ & $D_{M, \min}$ & / & / & / \\
~ & $E_{\min}$ (eV) & -123.4 & -123.4 & -123.4 \\
~ & $N_E$ & 1 & 1 & 1 \\
~ & $N_M$ & 0 & 0 & 0 \\
\hline
\multirow{9}{*}{2} & 
$I_{M, 0}$ & / & / & / \\
~ & $D_{M, 0}$ & / & / & / \\
~ & $I_{M, \min}$ & / & / & / \\
~ & $D_{M, \min}$ & / & / & / \\
~ & $E_{\min}$ (eV) & -123.4 & -123.4 & -123.4 \\
~ & $N_E$ & 1 & 2 & 5 \\
~ & $N_M$ & 0 & 0 & 0 \\
\hline
			\end{tabular}
		}
		\label{table:bench-lizro2}
	\end{table}
	
	\clearpage
	
	\begin{table}[h]
		\caption{Comparison on Li$_3$Ti$_3$Se$_6$O$_3$ benchmark crystal, whose estimated stable energy is -80.44 eV.}
		\renewcommand\arraystretch{1.5}
		\centering
		\resizebox{14cm}{!}{
			\begin{tabular}{ccccc}
\hline
\multirow{2}{*}{Seed} & \multirow{2}{*}{Metrics} & \multicolumn{3}{c}{Optimizer} \\
~ & ~ & Crystal param. + Hyperopt BO & CALYPSO & CRYSIM + Amplify \\
\hline
\multirow{9}{*}{0} & 
$I_{M, 0}$ & / & / & / \\
~ & $D_{M, 0}$ & / & / & / \\
~ & $I_{M, \min}$ & / & / & / \\
~ & $D_{M, \min}$ & / & / & / \\
~ & $E_{\min}$ (eV) & -88.96 & -89.3 & -89.14 \\
~ & $N_E$ & 1 & 1 & 1 \\
~ & $N_M$ & 0 & 0 & 0 \\
\hline
\multirow{9}{*}{1} & 
$I_{M, 0}$ & / & / & / \\
~ & $D_{M, 0}$ & / & / & / \\
~ & $I_{M, \min}$ & / & / & / \\
~ & $D_{M, \min}$ & / & / & / \\
~ & $E_{\min}$ (eV) & -88.78 & -89.3 & -89.06 \\
~ & $N_E$ & 1 & 1 & 1 \\
~ & $N_M$ & 0 & 0 & 0 \\
\hline
\multirow{9}{*}{2} & 
$I_{M, 0}$ & / & / & / \\
~ & $D_{M, 0}$ & / & / & / \\
~ & $I_{M, \min}$ & / & / & / \\
~ & $D_{M, \min}$ & / & / & / \\
~ & $E_{\min}$ (eV) & -89.36 & -89.32 & -89.12 \\
~ & $N_E$ & 1 & 1 & 1 \\
~ & $N_M$ & 0 & 0 & 0 \\
\hline
			\end{tabular}
		}
		\label{table:bench-litise2o}
	\end{table}
	
	\clearpage
	
	\begin{table}[h]
		\caption{Comparison on Li$_8$Zr$_4$O$_{12}$, whose estimated stable energy is -173.64 eV.}
		\renewcommand\arraystretch{1.5}
		\centering
		\resizebox{14cm}{!}{
			\begin{tabular}{ccccc}
\hline
\multirow{2}{*}{Seed} & \multirow{2}{*}{Metrics} & \multicolumn{3}{c}{Optimizer} \\
~ & ~ & Crystal param. + Hyperopt BO & CALYPSO & CRYSIM + Amplify \\
\hline
\multirow{9}{*}{0} & 
$I_{M, 0}$ & / & / & / \\
~ & $D_{M, 0}$ & / & / & / \\
~ & $I_{M, \min}$ & / & / & / \\
~ & $D_{M, \min}$ & / & / & / \\
~ & $E_{\min}$ (eV) & -173.51 & -172.36 & -173.65 \\
~ & $N_E$ & 1 & 1 & 2 \\
~ & $N_M$ & 0 & 0 & 0 \\
\hline
\multirow{9}{*}{1} & 
$I_{M, 0}$ & / & / & / \\
~ & $D_{M, 0}$ & / & / & / \\
~ & $I_{M, \min}$ & / & / & / \\
~ & $D_{M, \min}$ & / & / & / \\
~ & $E_{\min}$ (eV) & -170.96 & -173.64 & -173.64 \\
~ & $N_E$ & 1 & 1 & 1 \\
~ & $N_M$ & 0 & 0 & 0 \\
\hline
\multirow{9}{*}{2} & 
$I_{M, 0}$ & / & / & / \\
~ & $D_{M, 0}$ & / & / & / \\
~ & $I_{M, \min}$ & / & / & / \\
~ & $D_{M, \min}$ & / & / & / \\
~ & $E_{\min}$ (eV) & -172.95 & -172.57 & -173.32 \\
~ & $N_E$ & 1 & 1 & 1 \\
~ & $N_M$ & 0 & 0 & 0 \\
\hline
			\end{tabular}
		}
		\label{table:li8zr4o12}
	\end{table}
	
	\clearpage
	
	\begin{table}[h]
    %\vspace{5em}
	\caption{Lowest energies of structures discovered by different CSP optimizers, in which \textbf{bold} values are the lowest average energies achieved for each material system, and \underline{underlined} values indicate materials accordant with ground states, determined by \texttt{StructureMatcher} function in \texttt{pymatgen} package \cite{ONG2013314}. (unit: eV)}
	\renewcommand\arraystretch{1.5}
	\centering
	\resizebox{17cm}{!}{
		\begin{tabular}{cccccccc}
\hline
System & CSP optimizer & Seed 0 & Seed 1 & Seed 2 & Seed 3 & Seed 4 & Average \\
\hline

\multirow{5}{*}{Y$_{6}$Co$_{51}$} & RG & -390.02 & \underline{-396.32} & -394.02 & -391.34 & -393.89 & -393.12±2.21 \\
~ & CRYSPY RG & -392.2 & -392.09 & -395.5 & -396.31 & -392.56 & -393.73±1.8 \\
~ & Crystal param. + Hyperopt BO & -398.73 & \underline{-406.26} & -393.58 & \underline{-399.1} & -397.39 & \textbf{-399.01±4.12} \\
~ & CALYPSO & -396.31 & -394.55 & -395.28 & -399.37 & -394.36 & -395.97±1.83 \\
~ & CRYSIM + Amplify & -396.41 & -391.21 & -394.55 & -390.16 & -397.49 & -393.96±2.86 \\

\hline

\multirow{5}{*}{Ca$_{24}$Al$_{16}$(SiO$_{4}$)$_{24}$} & RG & -1025.94 & -1025.72 & -993.43 & -1040.35 & -1019.79 & -1021.05±15.38 \\
~ & CRYSPY RG & -1054.22 & -1073.31 & -1111.53 & -1057.77 & -1076.66 & -1074.7±20.34 \\
~ & Crystal param. + Hyperopt BO & -1066.44 & -1080.53 & -1073.36 & -1072.86 & -1033.41 & -1065.32±16.57 \\
~ & CALYPSO & -1101.4 & -1100.32 & -1104.03 & -1121.48 & -1091.78 & -1103.8±9.75 \\
~ & CRYSIM + Amplify & \underline{-1186.57} & \underline{-1194.67} & \underline{-1197.54} & \underline{-1197.59} & -1124.42 & \textbf{-1180.16±28.16} \\

\hline

\multirow{5}{*}{(SiO$_{2}$)$_{96}$} & RG & -1796.63 & -1805.58 & -1849.92 & -1848.66 & -1819.43 & -1824.04±21.86 \\    
~ & CRYSPY RG & -1914.3 & -1865.77 & -1878.39 & -1890.58 & -1859.76 & -1881.76±19.43 \\
~ & Crystal param. + Hyperopt BO & -1884.69 & -1789.12 & -1799.85 & -1757.42 & -1786.07 & -1803.43±42.99 \\
~ & CALYPSO & -2018.52 & -1996.8 & -2059.93 & -1982.44 & -2018.42 & -2015.22±26.21 \\
~ & CRYSIM + Amplify & -2272.18 & -2015.84 & -2076.38 & -2001.66 & -1869.37 & \textbf{-2047.09±131.26} \\

\hline
		\end{tabular}
	}
	\label{table:mainresults}
    \end{table}

    \clearpage
    
    	\begin{table}[h]
    	\caption{Comparison on Y$_6$Co$_{51}$, whose estimated stable energy is -406.26 eV.}
    	\renewcommand\arraystretch{1.2}
\centering
\resizebox{16cm}{!}{
    		\begin{tabular}{ccccccc}
\hline
\multirow{2}{*}{Seed} & \multirow{2}{*}{Metrics} & \multicolumn{5}{c}{Optimizer} \\
~ & ~ & RG & CRYSPY RG & Crystal param. + Hyperopt BO & CALYPSO & CRYSIM + Amplify \\
\hline
\multirow{9}{*}{0} & 
$I_{M, 0}$ & / & / & / & / & / \\
~ & $D_{M, 0}$ & / & / & / & / & / \\
~ & $I_{M, \min}$ & / & / & / & / & / \\
~ & $D_{M, \min}$ & / & / & / & / & / \\
~ & $E_{\min}$ (eV) & -390.02 & -392.2 & -398.73 & -396.31 & -396.41 \\
~ & $N_E$ & 1 & 1 & 1 & 2 & 1 \\
~ & $N_M$ & 0 & 0 & 0 & 0 & 0 \\
\hline
\multirow{9}{*}{1} & 
$I_{M, 0}$ & 68 & / & 250 & / & / \\
~ & $D_{M, 0}$ & 0.47193 & / & 0.00981 & / & / \\
~ & $I_{M, \min}$ & 68 & / & 250 & / & / \\
~ & $D_{M, \min}$ & 0.47193 & / & 0.00981 & / & / \\
~ & $E_{\min}$ (eV) & -396.32 & -392.09 & -406.26 & -394.55 & -391.21 \\
~ & $N_E$ & 1 & 1 & 1 & 1 & 1 \\
~ & $N_M$ & 1 & 0 & 1 & 0 & 0 \\
\hline
\multirow{9}{*}{2} & 
$I_{M, 0}$ & / & / & / & / & / \\
~ & $D_{M, 0}$ & / & / & / & / & / \\
~ & $I_{M, \min}$ & / & / & / & / & / \\
~ & $D_{M, \min}$ & / & / & / & / & / \\
~ & $E_{\min}$ (eV) & -394.02 & -395.5 & -393.58 & -395.28 & -394.55 \\
~ & $N_E$ & 1 & 1 & 1 & 1 & 1 \\
~ & $N_M$ & 0 & 0 & 0 & 0 & 0 \\
\hline
\multirow{9}{*}{3} & 
$I_{M, 0}$ & / & / & 14 & / & / \\
~ & $D_{M, 0}$ & / & / & 0.4382 & / & / \\
~ & $I_{M, \min}$ & / & / & 14 & / & / \\
~ & $D_{M, \min}$ & / & / & 0.4382 & / & / \\
~ & $E_{\min}$ (eV) & -391.34 & -396.31 & -399.1 & -399.37 & -390.16 \\
~ & $N_E$ & 1 & 1 & 1 & 1 & 1 \\
~ & $N_M$ & 0 & 0 & 1 & 0 & 0 \\
\hline
\multirow{9}{*}{4} & 
$I_{M, 0}$ & / & / & / & / & / \\
~ & $D_{M, 0}$ & / & / & / & / & / \\
~ & $I_{M, \min}$ & / & / & / & / & / \\
~ & $D_{M, \min}$ & / & / & / & / & / \\
~ & $E_{\min}$ (eV) & -393.89 & -392.56 & -397.39 & -394.36 & -397.49 \\
~ & $N_E$ & 1 & 1 & 1 & 1 & 1 \\
~ & $N_M$ & 0 & 0 & 0 & 0 & 0 \\
\hline
    		\end{tabular}
    	}
    	\label{table:y6co51}
    \end{table}
    
    \clearpage
    
        	\begin{table}[h]
    	\caption{Comparison on Ca$_{24}$Al$_{16}$(SiO$_4$)$_{24}$, whose estimated stable energy is -1197.59 eV.}
    	\renewcommand\arraystretch{1.2}
\centering
\resizebox{16cm}{!}{
    		\begin{tabular}{ccccccc}
\hline
\multirow{2}{*}{Seed} & \multirow{2}{*}{Metrics} & \multicolumn{5}{c}{Optimizer} \\
~ & ~ & RG & CRYSPY RG & Crystal param. + Hyperopt BO & CALYPSO & CRYSIM + Amplify \\
\hline
\multirow{9}{*}{0} & 
$I_{M, 0}$ & / & / & / & / & 212 \\
~ & $D_{M, 0}$ & / & / & / & / & 0.48172 \\
~ & $I_{M, \min}$ & / & / & / & / & 212 \\
~ & $D_{M, \min}$ & / & / & / & / & 0.48172 \\
~ & $E_{\min}$ (eV) & -1025.94 & -1054.22 & -1066.44 & -1101.4 & -1186.57 \\
~ & $N_E$ & 1 & 1 & 1 & 1 & 1 \\
~ & $N_M$ & 0 & 0 & 0 & 0 & 1 \\
\hline
\multirow{9}{*}{1} & 
$I_{M, 0}$ & / & / & / & / & 247 \\
~ & $D_{M, 0}$ & / & / & / & / & 0.03205 \\
~ & $I_{M, \min}$ & / & / & / & / & 247 \\
~ & $D_{M, \min}$ & / & / & / & / & 0.03205 \\
~ & $E_{\min}$ (eV) & -1025.72 & -1073.31 & -1080.53 & -1100.32 & -1194.67 \\
~ & $N_E$ & 1 & 1 & 1 & 1 & 1 \\
~ & $N_M$ & 0 & 0 & 0 & 0 & 1 \\
\hline
\multirow{9}{*}{2} & 
$I_{M, 0}$ & / & / & / & / & 247 \\
~ & $D_{M, 0}$ & / & / & / & / & 0.006 \\
~ & $I_{M, \min}$ & / & / & / & / & 247 \\
~ & $D_{M, \min}$ & / & / & / & / & 0.006 \\
~ & $E_{\min}$ (eV) & -993.43 & -1111.53 & -1073.36 & -1104.03 & -1197.54 \\
~ & $N_E$ & 1 & 1 & 1 & 1 & 1 \\
~ & $N_M$ & 0 & 0 & 0 & 0 & 1 \\
\hline
\multirow{9}{*}{3} & 
$I_{M, 0}$ & / & / & / & / & 279 \\
~ & $D_{M, 0}$ & / & / & / & / & 0.00127 \\
~ & $I_{M, \min}$ & / & / & / & / & 279 \\
~ & $D_{M, \min}$ & / & / & / & / & 0.00127 \\
~ & $E_{\min}$ (eV) & -1040.35 & -1057.77 & -1072.86 & -1121.48 & -1197.59 \\
~ & $N_E$ & 1 & 1 & 1 & 1 & 1 \\
~ & $N_M$ & 0 & 0 & 0 & 0 & 1 \\
\hline
\multirow{9}{*}{4} & 
$I_{M, 0}$ & / & / & / & / & / \\
~ & $D_{M, 0}$ & / & / & / & / & / \\
~ & $I_{M, \min}$ & / & / & / & / & / \\
~ & $D_{M, \min}$ & / & / & / & / & / \\
~ & $E_{\min}$ (eV) & -1019.79 & -1076.66 & -1033.41 & -1091.78 & -1124.42 \\
~ & $N_E$ & 1 & 1 & 1 & 1 & 1 \\
~ & $N_M$ & 0 & 0 & 0 & 0 & 0 \\
\hline
    		\end{tabular}
    	}
    	\label{table:ca24al16si24o96}
    \end{table}
    
    \clearpage
    
        	\begin{table}[h]
    	\caption{Comparison on (SiO$_2$)$_{96}$, whose estimated stable energy is -2272.57 eV.}
    	\renewcommand\arraystretch{1.2}
    	\centering
    	\resizebox{16cm}{!}{
    		\begin{tabular}{ccccccc}
\hline
\multirow{2}{*}{Seed} & \multirow{2}{*}{Metrics} & \multicolumn{5}{c}{Optimizer} \\
~ & ~ & RG & CRYSPY RG & Crystal param. + Hyperopt BO & CALYPSO & CRYSIM + Amplify \\
\hline
\multirow{9}{*}{0} & 
$I_{M, 0}$ & / & / & / & / & / \\
~ & $D_{M, 0}$ & / & / & / & / & / \\
~ & $I_{M, \min}$ & / & / & / & / & / \\
~ & $D_{M, \min}$ & / & / & / & / & / \\
~ & $E_{\min}$ (eV) & -1796.63 & -1914.3 & -1884.69 & -2018.52 & -2272.18 \\
~ & $N_E$ & 1 & 1 & 1 & 1 & 1 \\
~ & $N_M$ & 0 & 0 & 0 & 0 & 0 \\
\hline
\multirow{9}{*}{1} & 
$I_{M, 0}$ & / & / & / & / & / \\
~ & $D_{M, 0}$ & / & / & / & / & / \\
~ & $I_{M, \min}$ & / & / & / & / & / \\
~ & $D_{M, \min}$ & / & / & / & / & / \\
~ & $E_{\min}$ (eV) & -1805.58 & -1865.77 & -1789.12 & -1996.8 & -2015.84 \\
~ & $N_E$ & 1 & 1 & 1 & 1 & 1 \\
~ & $N_M$ & 0 & 0 & 0 & 0 & 0 \\
\hline
\multirow{9}{*}{2} & 
$I_{M, 0}$ & / & / & / & / & / \\
~ & $D_{M, 0}$ & / & / & / & / & / \\
~ & $I_{M, \min}$ & / & / & / & / & / \\
~ & $D_{M, \min}$ & / & / & / & / & / \\
~ & $E_{\min}$ (eV) & -1849.92 & -1878.39 & -1799.85 & -2059.93 & -2076.38 \\
~ & $N_E$ & 1 & 1 & 1 & 1 & 1 \\
~ & $N_M$ & 0 & 0 & 0 & 0 & 0 \\
\hline
\multirow{9}{*}{3} & 
$I_{M, 0}$ & / & / & / & / & / \\
~ & $D_{M, 0}$ & / & / & / & / & / \\
~ & $I_{M, \min}$ & / & / & / & / & / \\
~ & $D_{M, \min}$ & / & / & / & / & / \\
~ & $E_{\min}$ (eV) & -1848.66 & -1890.58 & -1757.42 & -1982.44 & -2001.66 \\
~ & $N_E$ & 1 & 1 & 1 & 1 & 1 \\
~ & $N_M$ & 0 & 0 & 0 & 0 & 0 \\
\hline
\multirow{9}{*}{4} & 
$I_{M, 0}$ & / & / & / & / & / \\
~ & $D_{M, 0}$ & / & / & / & / & / \\
~ & $I_{M, \min}$ & / & / & / & / & / \\
~ & $D_{M, \min}$ & / & / & / & / & / \\
~ & $E_{\min}$ (eV) & -1819.43 & -1859.76 & -1786.07 & -2018.42 & -1869.37 \\
~ & $N_E$ & 1 & 1 & 1 & 1 & 1 \\
~ & $N_M$ & 0 & 0 & 0 & 0 & 0 \\
\hline
    		\end{tabular}
    	}
    	\label{table:si96o192}
    \end{table}
    
    \clearpage
	
	\begin{table}[h]
		\caption{Numbers of filtered abnormal structures due to containing extremely close atom pairs among the 300 generations, averaged on five trials.}
		\renewcommand\arraystretch{1.5}
		\centering
		\resizebox{11cm}{!}{
			\begin{tabular}{ccc}
\hline
System & CSP optimizer & Filtered number \\
\hline

\hline

\multirow{5}{*}{Y$_{6}$Co$_{51}$} & RG & 11±1 \\
~ & CRYSPY RG & 0±0 \\
~ & Crystal param. + Hyperopt BO & 38±8 \\
~ & CALYPSO & 0±0 \\
~ & CRYSIM + Amplify & 1±0 \\

\hline

\multirow{5}{*}{Ca$_{24}$Al$_{16}$(SiO$_{4}$)$_{24}$} & RG & 16±3 \\
~ & CRYSPY RG & 0±0 \\
~ & Crystal param. + Hyperopt BO & 21±5 \\
~ & CALYPSO & 0±0 \\
~ & CRYSIM + Amplify & 7±1 \\

\hline

\multirow{5}{*}{(SiO$_{2}$)$_{96}$} & RG & 61±15 \\
~ & CRYSPY RG & 0±0 \\
~ & Crystal param. + Hyperopt BO & 54±7 \\
~ & CALYPSO & 0±0 \\
~ & CRYSIM + Amplify & 26±9 \\

\hline
			\end{tabular}
		}
		\label{table:filternumclassical}
	\end{table}
	
	\clearpage
	
		\begin{table}[h]
		\caption{Number of bits in resulting CRYSIM embeddings for different systems, composed of lattice parameters, symmetry information and atomic positions segments.}
		\renewcommand\arraystretch{1.5}
		\centering
		\resizebox{14cm}{!}{
			\begin{tabular}{cccccc}
				\hline
				\multirow{2}{*}{LDR} & \multicolumn{2}{c}{\multirow{2}{*}{Parameter}} & \multicolumn{3}{c}{System} \\
				~ & ~ & ~ & Y$_{6}$Co$_{51}$ & Ca$_{24}$Al$_{16}$(SiO$_{4}$)$_{24}$ & (SiO$_{2}$)$_{96}$ \\
				\hline
				\multirow{7}{*}{5 * 5 * 5} & \multirow{2}{*}{Lattice parameters} & Lattice length & 35 & 37 & 37 \\
				~ & ~ & Lattice angle & 40 & 40 & 40 \\
				~ & \multirow{3}{*}{Symmetry information} & Crystal system & 7 & 7 & 7 \\
				~ & ~ & Space group & 19 & 68 & 68 \\
				~ & ~ & Wyckoff positions combination & 300 & 300 & 300 \\
				~ & Atomic positions & / & 125 & 125 & 125 \\
				~ & Total & / & 801 & 1106 & 856 \\
				\hline
				\multirow{7}{*}{7 * 7 * 7} & \multirow{2}{*}{Lattice parameters} & Lattice length & 49 & 53 & 53 \\
				~ & ~ & Lattice angle & 40 & 40 & 40 \\
				~ & \multirow{3}{*}{Symmetry information} & Crystal system & 7 & 7 & 7 \\
				~ & ~ & Space group & 19 & 68 & 68 \\
				~ & ~ & Wyckoff positions combination & 300 & 300 & 300 \\
				~ & Atomic positions & / & 343 & 343 & 343 \\
				~ & Total & / & 1279 & 2026 & 1340 \\
				\hline
				\multirow{7}{*}{9 * 9 * 9} & \multirow{2}{*}{Lattice parameters} & Lattice length & 63 & 68 & 68 \\
				~ & ~ & Lattice angle & 40 & 40 & 40 \\
				~ & \multirow{3}{*}{Symmetry information} & Crystal system & 7 & 7 & 7 \\
				~ & ~ & Space group & 19 & 68 & 68 \\
				~ & ~ & Wyckoff positions combination & 300 & 300 & 300 \\
				~ & Atomic positions & / & 729 & 729 & 729 \\
				~ & Total & / & 2093 & 3615 & 2157 \\
				\hline
				\multirow{7}{*}{12 * 12 * 12} & \multirow{2}{*}{Lattice parameters} & Lattice length & 85 & 90 & 90 \\
				~ & ~ & Lattice angle & 40 & 40 & 40 \\
				~ & \multirow{3}{*}{Symmetry information} & Crystal system & 7 & 7 & 7 \\
				~ & ~ & Space group & 19 & 68 & 68 \\
				~ & ~ & Wyckoff positions combination & 300 & 300 & 300 \\
				~ & Atomic positions & / & 1728 & 1728 & 1728 \\
				~ & Total & / & 4157 & 7677 & 4221 \\
				\hline
				\multirow{7}{*}{15 * 15 * 15} & \multirow{2}{*}{Lattice parameters} & Lattice length & 106 & 113 & 113 \\
				~ & ~ & Lattice angle & 40 & 40 & 40 \\
				~ & \multirow{3}{*}{Symmetry information} & Crystal system & 7 & 7 & 7 \\
				~ & ~ & Space group & 19 & 68 & 68 \\
				~ & ~ & Wyckoff positions combination & 300 & 300 & 300 \\
				~ & Atomic positions & / & 3375 & 3375 & 3375 \\
				~ & Total & / & 7514 & 14334 & 7584 \\
				\hline
			\end{tabular}
		}
		\label{table:length}
	\end{table}
	
	\clearpage

	\begin{table}[h]
	\caption{Numbers of filtered structures for CRYSIM optimizers with (Y) and without (N) MID-related procedures due to containing extremely close atom pairs, in which \textbf{bold} values are the lower ones for each LDR. Each value is averaged on three seeds.}
	\renewcommand\arraystretch{1.5}
	\centering
	\resizebox{15cm}{!}{
		\begin{tabular}{ccccccc}
\hline
\multirow{2}{*}{System} & \multirow{2}{*}{MID proc.} & \multicolumn{5}{c}{Lattice discretization resolution} \\
~ & ~ & 5 * 5 * 5 & 7 * 7 * 7 & 9 * 9 * 9 & 12 * 12 * 12 & 15 * 15 * 15 \\
\hline
\multirow{2}{*}{Y$_{6}$Co$_{51}$} & N & 29±15 & 23±4 & 20±12 & 20±10 & 8±0 \\
~ & Y & \textbf{2±1} & \textbf{2±0} & \textbf{4±2} & \textbf{1±0} & \textbf{1±0} \\
\hline
\multirow{2}{*}{Ca$_{24}$Al$_{16}$(SiO$_{4}$)$_{24}$} & N & 26±12 & 19±2 & 38±9 & 42±19 & 38±18 \\
~ & Y & \textbf{8±4} & \textbf{8±2} & \textbf{11±6} & \textbf{7±1} & \textbf{4±0} \\
\hline
\multirow{2}{*}{(SiO$_{2}$)$_{96}$} & N & / & 58±13 & 62±19 & 70±29 & 68±10 \\
~ & Y & / & \textbf{21±11} & \textbf{28±12} & \textbf{22±7} & \textbf{26±4} \\
\hline
		\end{tabular}
	}
	\label{table:as-filternumpp}
	\end{table}	
	
	% \section{Explanations about Abnormal Generated Structures}
	% Due to limitation of machine learning potential (MLP) adopted in this work, estimated energies of some abnormal structures, especially those in which atoms are extremely close to each other, might be unreasonably low. When exhibiting results, those materials are ignored by a pre-set energy threshold, which means that materials with energy below the threshold will not be considered. Taking (SiO$_2$)$_{96}$ system as an example, this section provides explanations about how to determine the threshold.
		\clearpage
		
	\section{Application of Wyckoff Positions Combinations Lists in CRYSIM}
	\label{sec:wpc}
	
\subsection{Constructing Crystal Structures based on Wyckoff Positions}

In the vast potential energy surface (PES), most of the energetically stable crystals existing in the nature should have symmetry \cite{RN1158}, as is proved theoretically \cite{WALES1998330}. Therefore, when constructing crystal structures, such as recovering configurations from powder diffraction data \cite{Deng:ce5104} or crystal structure prediction (CSP) from unit cell compositions \cite{RN957}, symmetry is considered to increase the possibility of deriving reasonable crystals. 
Symmetry of three-dimensional crystals are depicted by 230 space groups (SGs), each of which is formed by a set of symmetry operations $P_{SG}$. A crystal which is symmetric with respect to an SG does not change under the corresponding operations.
That is to say, on the one hand, for each atom in the configuration, all possible positions that can be reached by conducting any operations in $P_{SG}$ on it have been occupied by other atoms of the same species simultaneously.
On the other hand, given an SG $S$, we can define a set of positions, each of which only contain 3D points that are equivalent with respect to the SG \cite{Julian2024}, i.e., $\mathbb{W}_{0, S}=\{W_{0, S, i}| \forall O\in S, \forall P\in W_{0, S, i}, O\cdot P \in W_{0, S, i}\}$.
These positions, each may contain more than one Cartesian coordinates, are called Wyckoff positions (WPs).
%The positions designated by an SG can be divided into sets of Wyckoff positions (WPs) of different site symmetries with operations $P_{s, SG} \subseteq P_{SG}$, which means that each single point in the WPs remain static under operations in $P_{s, SG}$. 
For instance, WPs of the SG $P321$ consist of
\begin{equation}
	\left\{
	\begin{aligned}
		W_{0, P321, 1}=&\{(0,0,0)\}, \\
		W_{0, P321, 2}=&\{(0,0,1/2)\}, \\
		W_{0, P321, 3}=&\{(0,0,z), (0,0,-z)\}, \\
		W_{0, P321, 4}=&\{(1/3, 2/3, z), (2/3, 1/3, -z)\}, \\
		W_{0, P321, 5}=&\{(x,0,0), (0,x,0), (-x,-x,0)\}, \\
		W_{0, P321, 6}=&\{(x,0,1/2), (0,x,1/2), (-x,-x,1/2)\}, \\
		W_{0, P321, 7}=&\{(x,y,z), (-y,x-y,z), (-x+y,-x,z), (y,x,-z), (x-y,-y,-z), (-x,-x+y,-z)\}, 
	\end{aligned}
	\right.
	\label{wp1}
\end{equation}
so that each $W_{0, P321, i}, i=1, \ldots, 7,$ does not change under any symmetry operations in $P321$. The size of $W_{0, P321, i}$, i.e., its multiplicity, indicates the number of points that should be involved to fulfill the symmetry. Besides, for one SG, the WP with the largest multiplicity ($W_{0, P321, 7}$ for $P321$) is called the general position, and other WPs are special positions \cite{Julian2024}.

Accordingly, when constructing a crystal structure, based on the stoichiometry, its symmetry can be implemented by only adding atoms to variables of WPs. In this process, three basic rules should be observed. First, if one WP is selected, all coordinates in the WP should be included in the configuration, otherwise the symmetry of the WP is not maintained. Second, atoms in one WP should have the same element species. Third, the summation of multiplicity of all used WPs for one element should be equal to the number of atoms of that element in the unit cell, otherwise the system is not constructed.
Based on that, there will be multiple ways to combine WPs for constructing material systems given a specific chemical composition, which are defined as WPs combinations (WPCs) in this work. Additionally, we note that WPs in formulas \ref{wp1}, \ref{wpbasic} and \ref{wp2} are labeled as $W_0$, but WPCs are as $W$ for differentiation.
Taking the A$_4$B$_6$ system as an example, if four A and six B atoms in a configuration satisfying the following relationship denoted by either of the WPCs: 
\begin{equation}
	\left\{
	\begin{aligned}
		A_1&: (1/3, 2/3, z_1),\\
		A_2&: (2/3, 1/3, -z_1),\\
		A_3&: (1/3, 2/3, z_2),\\
		A_4&: (2/3, 1/3, -z_2),\\
		B_1&: (x_3, y_3 ,z_3), \\
		B_2&: (-y_3, x_3-y_3, z_3), \\
		B_3&: (-x_3+y_3, -x_3, z_3), \\
		B_4&: (y_3, x_3, -z_3), \\
		B_5&: (x_3-y_3, -y_3, -z_3), \\
		B_6&: (-x_3, -x_3+y_3, -z_3),
	\end{aligned}
	\right.
\end{equation}
or 
\begin{equation}
	\left\{
	\begin{aligned}
		A_1&: (0, 0, 1/2),\\
		A_2&: (x_1, 0, 1/2), \\
		A_3&: (0, x_1, 1/2), \\
		A_4&: (-x_1, -x_1, 1/2), \\
		B_1&: (x_2, 0, 1/2), \\
		B_2&: (0,x_2,1/2), \\
		B_3&: (-x_2,-x_2,1/2), \\
		B_4&: (x_3, 0, 0), \\
		B_5&: (0, x_3, 0), \\
		B_6&: (-x_3, -x_3, 0),
	\end{aligned}
	\right.
\end{equation}
the structure has symmetry defined by the $P321$ SG. In total, there are 121 possible WPCs for A$_4$B$_6$ for the SG.

In general, suppose that we hope to build a material configuration of a system $A_{1, a_1}A_{2, a_2}\ldots A_{m, a_m}$, which contains $m$ elements and $a_i$ atoms for the $i$-th element in the unit cell, and we require the configuration to have the symmetry of SG $S$.
Then, WPs are combined, 
%in which $W_{0, S, k}$ is adopted $x_k$ times, 
leading to a set of WPCs $\mathbb{W}_S$, so that for $W_S\in \mathbb{W}_S$ the summations of multiplicity of WPs employed for each element are equal to their frequency in the unit cell.
This relationship can be formulated as
\begin{equation}
	\begin{aligned}
		\mathbb{W}_S = \{&x_1 * W_{0, S, 1} + x_2 * W_{0, S, 2} + \ldots + x_{N_S} * W_{0, S, N_S}, x_1, \ldots, x_{N_S}\in \mathbb{N}| \\
		&\forall 1\leq i \leq m, \exists y_{i, 1}, \ldots, y_{i, N_S} \in \mathbb{N}, \\
		&(y_{i, 1} * |W_{0, S, 1}| + \ldots + y_{i, N_S} * |W_{0, S, N_{S}}| = a_i) \land
		(\forall 1\leq j \leq N_S,
		\sum_{i=1}^m y_{i, j} = x_j)
		\}
	\end{aligned}
	\label{wpbasic}
\end{equation}
where $N_S=|\mathbb{W}_{0, S}|$, and $|W_{0, S, i}|$ denotes the multiplicity of the $i$-th WP. This "+" operation between two WPs in this formula represents concatenation, which appends all 3D points of the second WP to the first one, resulting in a combination between them.
If all atoms of the same element type in a structure occupy coordinates designated by a set of WPs completely ($\sum_{j=1}^{N_S} y_{i, j} * W_{0, S, j}$), the structure can have symmetry of $S$.

\begin{figure}[htbp]
	\centering
	\includegraphics[width=\linewidth, center]{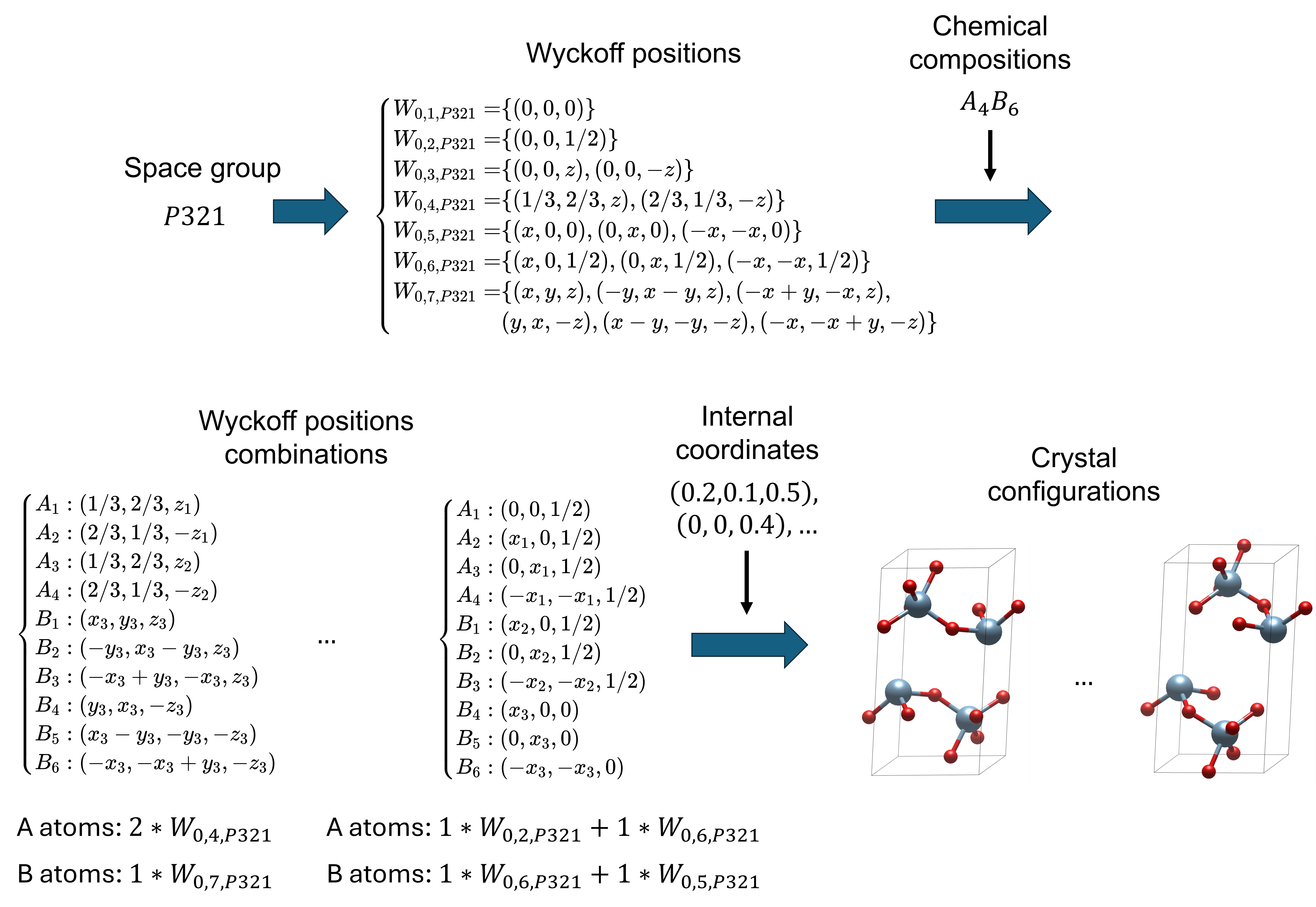}
	\caption{The workflow of building symmetric crystals utilizing WPCs lists, taking the SG $P321$ and system A$_4$B$_6$ as an example. First, WPs are decided for an SG, which can be found in public resources \cite{ITA6}. For a specific stoichiometry information, a list of WPCs is generated, ensuring the number of 3D points are accordant with frequency of atoms each element in the unit cell. This "+" operation is among WPs in this formula represents concatenation. For each WPC, after filling into the variables with independent coordinates, a structure having symmetry SG is derived. We note that the configurations shown in this Figure are illustrative, since they actually have different SGs.}
	\label{fig:wpbasics}
\end{figure}

In implementations, atom coordinates are first derived, and then inserted into the sites denoted by WPs. In this work, we define the coordinates before and after insertion as "independent coordinates" or "internal coordinates", and "external coordinates" for differentiation. The following \textbf{Fig. \ref{fig:wpbasics}} illustrates the process.

There are two main approaches in modern CSP software of incorporating WPs to generate symmetric structures \cite{RN957}.
In the first approach, starting from WPs with the highest multiplicity, i.e., general WPs, atoms (derived independent coordinates) are placed into WPs individually, and two atoms are merged if their distance are too close, leading to special WPs \cite{LYAKHOV20131172, RN957, RN958}.
In another approach, a set of all possible WPCs are calculated based on chemical composition of the system to be explored before structure generation. The crystal is constructed with one specific WPC directly after all independent coordinates are well prepared \cite{RN951, AVERY2019274, RN678}.
Since CRYSIM encodes and optimizes WPC indices, the second approach is adopted. As far as we are concerned, SMEPOC \cite{Deng:db5062} devoted the first effort to deduct a complete WPC list from a unit cell composition. Besides, RandSpg \cite{AVERY2017208} and GN-OA \cite{RN678} provided open-source code for the same target in a recursive workflow.

In CRYSIM, the implementation from GN-OA is integrated. 
In the algorithms, WPs are first gathered for each element in the system according to stoichiometry, and then combined by Cartesian product among the WPs sets.
Compared with the original code, WPs of an SG are sorted in descending orders based on multiplicity, ensuring that WPCs not containing static coordinates are first generated. This modification is designed to increase success rate in the first several trials by preventing that the same coordinates appear in WPs of different elements.
There is also evidence that most crystals in the nature tend to occupy more general WPs \cite{RN957}.
To increase the process of WPCs calculation especially for large systems, the maximum number WPCs for each element is set as $10^5$. The combining process ends if $10^6$ WPCs are collected, or after one successful generation, $10^8$ trials continuously fail for one SG.

\subsection{Crystal Systems and Space Groups Compatible with Stoichiometry}

The chemical composition in unit cells of a material system limits the types of symmetry it can achieve. By computing the list of WPCs for all SGs, respectively, SGs compatible with the stoichiometry are defined as the ones for which at least one WPC can be used to build configurations.
For instance, there is no possible WPCs for any A$_4$B$_4$ systems given the $F4_132$ SG, thus this type of crystals can never have the symmetry. Furthermore, compatibility of CSs relies on SGs. In A$_3$B$_3$C$_3$ systems, none of SG numbers within $[195, 230]$ can be achieved, so that this materials family cannot have cubic lattices.
As an illustration, WPs of $P23$, the No.195 SG, are given as
\begin{equation}
	\left\{
	\begin{aligned}
		W_{0, P23, 1}=&\{(0,0,0)\}, \\
		W_{0, P23, 2}=&\{(1/2,1/2,1/2)\}, \\
		W_{0, P23, 3}=&\{(0, 1/2, 1/2), (1/2, 0, 1/2), (1/2, 1/2, 0)\}, \\
		W_{0, P23, 4,}=&\{(1/2, 0, 0), (0, 1/2, 0), (0, 0, 1/2)\}, \\
		W_{0, P23, 5}=&\{(x, x, x), (-x, -x, x), (-x, x, -x), (x, -x, -x)\}, \\
		W_{0, P23, 6}=&\{(x, 0, 0), (-x, 0, 0), (0, x, 0), (0, -x, 0), (0, 0, x), (0, 0, -x)\}, \\
		W_{0, P23, 7}=&\{(x, 0, 1/2), (-x, 0, 1/2), (1/2, x, 0), (1/2, -x, 0), (0, 1/2, x), (0, 1/2, -x)\}, \\
		W_{0, P23, 8}=&\{(x, 1/2, 0), (-x, 1/2, 0), (0, x, 1/2), (0, -x, 1/2), (1/2, 0, x), (1/2, 0, -x)\}, \\
		W_{0, P23, 9}=&\{(x, 1/2, 1/2), (-x, 1/2, 1/2), (1/2, x, 1/2), (1/2, -x, 1/2), (1/2, 1/2, x), (1/2, 1/2, -x)\}, \\
		W_{0, P23, 10}=&\{(x, y, z), (-x, -y, z), (-x, y, -z), (x, -y, -z), (z, x, y), (z, -x, -y), (-z, -x, y), 
		\\ &(-z, x, -y), (y, z, x), (-y, z, -x), (y, -z, -x), (-y, -z, x)\},
	\end{aligned}
	\right.
	\label{wp2}
\end{equation}
with multiplicities being 1, 1, 3, 3, 4, 6, 6, 6, 6, 12. Therefore, it is impossible to combine the WPs so that three A, three B and three C atoms can occupy at the same time, making $P23$ incompatible for A$_3$B$_3$C$_3$ systems.

Suppose the WPCs list is denoted as $\mathbb{W}_{all} = \{\mathbb{W}_S| S\in \mathbb{S}_{all}\}$, in which $\mathbb{S}_{all}$ includes all SGs from No.2 to 230. Some of WPC sets are empty for the specific stoichiometry, leading to
\begin{equation}
	\left\{
	\begin{aligned}
		\mathbb{W}_-&=\{\mathbb{W}_{S}||\mathbb{W}_{S}|=0\},\\
		\mathbb{W}_+&=\{\mathbb{W}_{S}||\mathbb{W}_{S}|\neq 0\},\\
		\mathbb{W}_{all}&=\mathbb{W}_- \cup \mathbb{W}_+.
	\end{aligned}
	\right.
\end{equation}
Then, the sets of SGs and CSs can be divided into 
\begin{equation}
	\left\{
	\begin{aligned}
		\mathbb{S}_-&=\{S|\mathbb{W}_{S} \in \mathbb{W}_-\},\\
		\mathbb{S}_+&=\{S|\mathbb{W}_{S} \in \mathbb{W}_+\},\\
		\mathbb{S}_{all}&=\mathbb{S}_- \cup \mathbb{S}_+,\\
		\mathbb{C}_-&=\{C|\mathbb{W}_{S} \in \mathbb{W}_-, \forall S \in \mathbb{S}_C\},\\
		\mathbb{C}_+&=\{C|\mathbb{W}_{S} \in \mathbb{W}_+, \exists S \in \mathbb{S}_C\},\\
		\mathbb{C}_{all}&=\mathbb{C}_- \cup \mathbb{C}_+,
	\end{aligned}
	\right.
\end{equation}
depending on compatibility, in which $\mathbb{S}_C$ denotes the set of SGs for a CS $C$, such as $\mathbb{S}_{cubic} = \{P23, F23, \ldots, Ia\overline{3}d\}$.
When constructing crystals by sampling SGs, it is necessary to exclude the incompatible ones beforehand to make the whole process robust.
		
		\clearpage
		
	\section{Considerations on the Design of CRYSIM Embeddings}
	\label{sec:discuss}
	
	\subsection{Simultaneously Solving Symmetry and Coordinates}
	
	 In an end-to-end CSP process, symmetry information is not included as prior knowledge, which, however, decides degrees of freedom of crystal parameters needed to be optimized. Since CRYSIM optimizes symmetry, lattice parameters and atomic coordinates simultaneously, it has to consider all possible parameters to prevent failure in decoding.
	For instance, the cubic lattice only requires 1 lattice length to be determined, but all the three lengths and three angles are still optimized by CRYSIM in case of the triclinic lattice finally being selected by the solver.
	
	Similar strategy is used for the number of independent atomic coordinates, which will always be consistent with stoichiometry of the input system. Different from the situation for lattice parameters, all coordinates of one element species are optimized inside one discrete lattice, instead of owning a specific vector segment respectively.
		In CRYSIM, the bits lying on the leftmost side are selected, as is discussed in \textbf{Method} section. This may introduce a bias that the obtained independent coordinates usually tend to appear in certain area of the lattice.
		We expect that it does not significantly influence crystal construction, since after being placed into WPs, the external coordinates always uniformly distribute in the lattice.
		Besides, the embeddings of the optimization problem are highly sparse, allowing the Ising solver to explore the solutions whose leftmost 1-bits are located in the right side of the vector segment.
	
	\subsection{Priority in Deciding Symmetry}
	WPCs define the direct rules that determine configurations. An option of encoding symmetry is to only encode the WPC index, and optimize it directly.
		If all WPCs for all SGs are listed and indexed, i.e., instead of independently indexed for each SG, as designed in CRYSIM, corresponding SG, as well as CS, can be determined by the index of obtained WPC.
		Nevertheless, this strategy can lead to unbalanced representation on SGs. Some SGs may be related to millions of solutions, while others may be hundred. But in order to find stable structures, a correct SG is of great importance.
		Similarly, CS has a even higher priority than SG. According to statistics of systems in MP \cite{LI2021110686}, half of the stable materials having multiple isomers still share the same CS, which is the prerequisite of correctly predicting the structure from composition. We try to prevent that some CSs have a higher possibility to be selected since they include more SGs, though the possibility of selecting SGs, based on this encoding approach, can be different, as is shown by an example in \textbf{Method} section.
	
\begin{figure}[htbp]
		\centering
		\includegraphics[width=5cm, center]{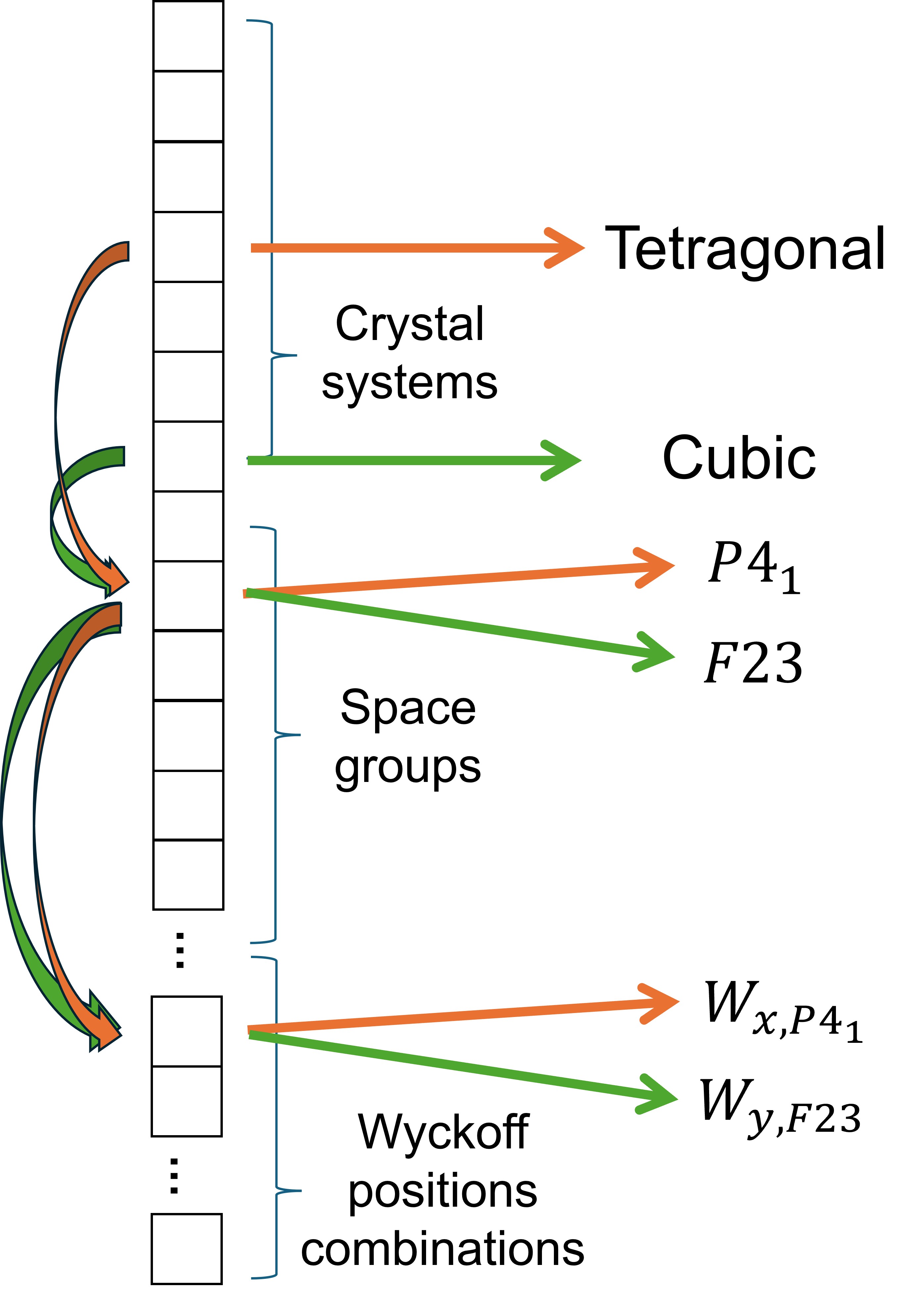}
		\caption{
			%Illustration of the encoding and decoding process. \textbf{a} Corresponding elements to be encoded and decoded from a vector with length 18 or 4 bits, longer or shorter than the embeddings of 10, respectively, calculated following equations \ref{M-eq:onehotsg}, \ref{M-eq:onehotwy} and \ref{M-eq:onehotde} in the main manuscript. The dashed lines indicate that the input one-hot vectors can not be recovered in decoding. \textbf{b}
			Orders of determination of symmetry information: from CS to SG, and then to WPC. For the same bits in SG and WP combination segments, different solutions can be decoded, dependent upon the decoded CS. The SG and WPC denoted by orange arrows are decoded when the solved CS is tetragonal, and green ones are for cubic lattice.}
		\label{fig:emb}
	\end{figure}

	Accordingly, there exists a sequence of symmetry information determination, as illustrated in \textbf{Fig. \ref{fig:emb}}, which means that what one bit represents in the SG segment is decided by the solved CS, and WPC is decided by the solved SG. In the SG segment, two solutions may have 1-bit at the same location, but if they have different CS bits, they have different SG after decoding.
	
	\clearpage
		
	\section{Details of Training Regression Models and Ising Solver Hyperarameters}
	\label{sec:ising}
	
	\subsection{Training Factorization Machine Models}
	
	In this work, 2-order Factorization Machine (FM) is implemented based on formula \ref{eq:fm2} \cite{5694074} in the main manuscript using \texttt{PyTorch} package \cite{NEURIPS2019_bdbca288}.
	Before training, energies in the derived dataset is standardized to 0 mean and 1 standard deviation.
	Then, the dataset from random generation is split into 9:1 as training set and validation set, and the trained model is assessed on validation set during training. If the validation loss does not decrease continuously for \texttt{patience} epochs, the training process will finish and the trainable parameters are used to build the objective function. In implementation, the value of \texttt{patience} is equal to half of the pre-set \texttt{max\_epoch} for training.
	
	Adam optimizer \cite{kingma2017adammethodstochasticoptimization} is applied to train the model, and training hyperparameters are fine-tuned using TPESampler in \texttt{optuna} package \cite{10.1145/3292500.3330701} with a grid-search manner from following ranges:
	
	\begin{table}[H]
		\renewcommand\arraystretch{1.5}
		\centering
		\resizebox{14cm}{!}{
			\begin{tabular}{cc}
				\hline
				Hyperparameter &  Fine-tuned range \\
				\hline
				max epoch & [300, 500, 800, 1000, 1500] \\
				batch size & [5, 10, 20, 50, 100] \\
				$K$ & [8, 16, 24, 32, 64] \\
				initial learning rate & [10$^{-2}$, 10$^{-3}$, 10$^{-4}$, 10$^{-5}$, 10$^{-6}$, 10$^{-7}$] \\
				\multirow{2}{*}{learning rate decay scheme} & ["None", "ReduceLROnPlateau", "LinearLR", \\
				~ & "ExponentialLR", "MultiStepLR", "SequentialLR"] \\
				weight decay for Adam & [10$^{-5}$, 10$^{-6}$, 10$^{-7}$, 10$^{-8}$, 10$^{-9}$] \\
				weight of RMSE and PCC in loss function & [(0, 1), (1, 1), (1, 10), (1, 100),
				(100, 1), (10, 1), (1, 0)] \\
				warming up steps & [0, 200, 500, 1000] \\
				EMA momentum & [1, 0.5, 0.1, 0.01] \\
				\hline
			\end{tabular}
		}
	\end{table}
	
	In this work, learning rate decay ends until the learning rate is equal to 10$^{-4}$ times of its initial value, which is implemented based \texttt{PyTorch} as follows.
	In "ReduceLROnPlateau", the learning rate times 0.9 (\texttt{factor=0.9}) if the validation loss does not decrease for 10 epochs (\texttt{patience=10}).
	In "LinearLR", the learning rate gradually decreases from the initial value (\texttt{start\_factor=1}) to the end value (\texttt{end\_factor=end\_lr / start\_lr}) within the first \texttt{0.8 * max\_epoch} epochs (\texttt{total\_iters=int(0.8 * max\_epoch)}) and remains unchanged.
	In "ExponentialLR", the learning rate decreases by multiplying 0.99 in each epoch (\texttt{gamma=0.99}), and stops with the end value.
	In "MultiStepLR", the learning rate decreases 4 times uniformly throughout the training duration, in which each time the learning rate times 0.1 (\texttt{gamma=0.1}).
	In "SequentialLR", the "LinearLR" strategy first performs \texttt{milestones} epochs (\texttt{total\_iters=int(milestones * 0.8)}). Then the learning rate is assigned back to the initial value, and "ExponentialLR" is applied with \texttt{gamma=0.993} until the end of training. The milestone is defined as \texttt{int(max\_epoch * 0.4)}.
	
	Besides, if \texttt{warming\_up\_steps > 0}, the learning rate will linearly increase from \texttt{start\_lr / warming\_up\_steps} to \texttt{start\_lr} in \texttt{warming\_up\_steps} steps (not epoch), and then decay starts from this epoch.
	
	The loss function is composed of root mean square error (RMSE) term and Pearson correlation coefficient (PCC) term, and the weight of the two terms is tuned with categories shown in "weight of RMSE and PCC in loss function" row, respectively. For instance, in "(1, 100)" category, the loss function is computed by
	\begin{align}
		loss = RMSE - 100 * PCC.
	\end{align}
	
	The EMA momentum is implemented by mixing trainable parameters of the last epoch with the ones in this epoch. The value represents the weight of new parameters, so that the procedure is not conducted if \texttt{EMA\_momentum = 1}.
	
	The set of hyperparameters that achieve the highest PCC are adopted for further experiments in this work, as summarized in the following table:
	
	\begin{table}[H]
		\renewcommand\arraystretch{1.5}
		\centering
		\resizebox{9.5cm}{!}{
			\begin{tabular}{cc}
				\hline
				Hyperparameter &  Value \\
				\hline
				max epoch & 800 \\
				batch size & 100 \\
				$K$ & 16 \\
				initial learning rate & 10$^{-3}$ \\
				learning rate decay scheme & "MultiStepLR" \\
				weight decay for Adam & 10$^{-9}$ \\
				weight of RMSE and PCC in loss function & (10, 1) \\
				warming up steps & 1000 \\
				EMA momentum & 1 \\
				\hline
			\end{tabular}
		}
	\end{table}
	
	Though there are many combinations to consider, the fine-tuning process does not cost a long time, since FM contains only tens of thousands of parameters.
	
	\subsection{Hyperarameters for Amplify as the Ising Solver}
    For Amplify \cite{RN1120}, \texttt{client.parameters.timeout} controls the annealing time in each solving process. In our implementation, its value depends on the number of bits in the objective function:
	
	\begin{table}[H]
		\renewcommand\arraystretch{1.5}
		\centering
		\resizebox{6cm}{!}{
			\begin{tabular}{cc}
				\hline
				Number of bits $x$ &  Timeout (ms) \\
				\hline
				$x <= 5000 $ &  30000 \\
				$5000 < x <= 8000 $ &  50000 \\
				$x > 8000 $ & 80000 \\
				\hline
			\end{tabular}
		}
	\end{table}
	
	In some cases, for objective functions of the same size, a smaller \texttt{timeout} may lead to a better performance than a larger one. However, we do not focus on tuning the parameter, but try to balance the performance and solving time needed.
	
	\clearpage
	
	\section{Hyperarameters of Classical CSP Algorithms}
	\label{sec:hp}
	
	For RG in CRYSPY \cite{RN958} and CALYPSO \cite{RN951}, interatomic distances matrices are given to restrict generated structures.
	For Y$_6$Co$_{51}$ and Ca$_{24}$Al$_{16}$(SiO$_4$)$_{24}$ systems,  minimum bond lengths for all atom pairs are 1.5 \AA \ , and for (SiO$_2$)$_{96}$ the distances are 1 \AA \ , respectively, to balance the stability of generated materials and difficulty of generation. 
	For ScBe$_5$, Ca$_4$S$_4$, Ba$_3$Na$_3$Bi$_3$, Li$_4$Zr$_4$O$_8$ and Li$_3$Ti$_3$Se$_6$O$_3$ systems, all minimal bond lengths are set as 1.5 \AA \  in CALYPSO.
	
	Bayesian optimization (BO) leveraged in this work is based on Tree-structured Parzen Estimator (TPE) models \cite{NIPS2011_86e8f7ab}, implemented using the hyperopt package  \cite{osti_10022769, RN678}. For parameters, \texttt{max\_evals} is set as 300, and structure relaxation is conducted for crystals generated in each trial, leading to 300 generations. Lower and upper bounds for lattice lengths and angles are the same as the ones in CRYSIM embeddings.
	
	For CRYSPY, only random generation (RG) is tested in this work, since optimization methods are encapsulated with calculation software, and pretrained universal machine learning potential, which is applied for energy estimation in this work, cannot be used. Apart from pair-wise distance matrices described in the main manuscript, parameters of RG are given as follows for all systems:
	
	\begin{table}[H]
	\renewcommand\arraystretch{1.5}
	\centering
	\resizebox{4cm}{!}{
		\begin{tabular}{cc}
			\hline
			Hyperparameter &  Value \\
			\hline
			nstage & 1 \\
			njob & 5 \\
			\hline
		\end{tabular}
	}
	\end{table}
	
	Other parameters, including range of space group numbers, are not indicated.
	
	For CALYPSO, apart from pair-wise distance matrices described in the main manuscript, parameters are given as follows for all systems:
	
	\begin{table}[H]
		\renewcommand\arraystretch{1.5}
		\centering
		\resizebox{5cm}{!}{
			\begin{tabular}{cc}
				\hline
				Hyperparameter &  Value \\
				\hline
				Ialgo & 2 \\
				PsoRatio & 0.6 \\
				PopSize & 10 \\
				NumberOfLbest & 4 \\
				NumberOfLocalOptim & 1 \\
				\hline
			\end{tabular}
		}
	\end{table}
	
	In all tests in this work, 30 iterations are conducted for CALYPSO to keep the total times of structure relaxation \texttt{PopSize * NumberOfLocalOptim * n\_iteration} equal to 300. Except for \texttt{PopSize} and \texttt{NumberOfLocalOptim}, parameters listed in the table are based on recommendation in the manual. Other parameters are not indicated in the input files.
	
	\clearpage

\end{document}